\documentclass[12pt]{article}
\usepackage[
top = 2.5cm, 
bottom = 2.5cm, 
left = 2.5cm, 
right = 2.5cm]{geometry}

\usepackage{jheppub}
\usepackage[sort&compress]{natbib}
\usepackage[utf8]{inputenc}
\usepackage{ifpdf}
\usepackage{braket}
\usepackage{graphicx}
\usepackage{caption}
\usepackage{subcaption}
\usepackage{slashed}
\usepackage{feynmf}
\usepackage{epstopdf}
\usepackage{amsmath, amssymb,url,bm,textgreek,upgreek}
\usepackage[svgnames]{xcolor}
\usepackage{hyperref}
\usepackage{dsfont}
\usepackage{float}

\def\tr{\operatorname{tr}}

\def\Vol{\operatorname{Vol}}

\def\tr{\operatorname{tr}}

\def\d{{\rm d}}

\def\Li{\operatorname{Li}}

\def\bryOPE{b}
\def\bulkOPE{c}

\def\tr{{\rm tr}}
\def\d{{\rm d}}

\def\CD{{\cal D}}

\def\CF{{\cal F}}
\def\CG{{\cal G}}

\def\CO{{\cal O}}

\title{The $O(N)$ model with $\phi^6$ potential in ${\mathbb R}^2 \times {\mathbb R}^+$}
\author{Christopher P. Herzog$^{a}$}
\author{and Nozomu Kobayashi$^{b,c}$}

\affiliation{
$^a$ Mathematics Department, King's College London, \\
The Strand, London,  WC2R 2LS, UK \\
$^b$ Kavli Institute for the Physics and Mathematics of the Universe (WPI), \\
The University of Tokyo Institutes for Advanced Study, The University of Tokyo, \\
Kashiwa, Chiba 277-8583, Japan \\
$^c$ Department of Physics, Faculty of Science,
The University of Tokyo,\\
Bunkyo-ku, Tokyo 113-0033, Japan \\ 
Emails: \href{mailto:christopher.herzog@kcl.ac.uk}{christopher.herzog@kcl.ac.uk},
\href{mailto:nozomu.kobayashi@ipmu.jp}{nozomu.kobayashi@ipmu.jp}.
}

\abstract{
We study the large $N$ limit of 
$O(N)$ scalar field theory with classically marginal $\phi^6$ interaction in 
three dimensions in the presence of a planar boundary.  
This theory has an approximate conformal invariance at large $N$.  
We find different phases of the theory corresponding to different boundary conditions for the 
scalar field.  Computing a one loop effective potential,
we examine the stability of these different phases.  The potential also allows us to determine a
boundary anomaly coefficient in the trace of the stress tensor.
We further 
compute the current and stress-tensor two point functions for the Dirichlet case 
and decompose them into boundary and bulk
conformal blocks.  The boundary limit of the stress tensor two point function allows us to compute the other boundary 
anomaly coefficient.  Both anomaly coefficients depend on the approximately marginal $\phi^6$ coupling.}

\preprint{IPMU20-0056}

\begin{document}
\maketitle

\section{Introduction}

Quantum field theory in the presence of a boundary has a long if little known history.  Important work
was done in the late seventies and early eighties in the context of surface critical phenomena.  A substantial
fraction of this work concerns the $O(N)$ scalar field theory 
with a $\phi^4$ interaction in the bulk and a relevant $\phi^2$ interaction on the boundary,
both in $4 - \epsilon$ dimensions and also in the limit of large $N$.  
Among other triumphs, estimates for 
surface critical exponents  were obtained and successfully matched with experimental data in some
instances.  (See e.g.\ \cite{PTandCPreviews,binder1983critical,Diehl:1996kd} for reviews.)  
Literature on 
$\phi^6$ theory in three dimensions with boundary, however, is scarce.  
A mean field analysis along with an expansion in $3-\epsilon$ dimensions can be found in refs.\
\cite{BinderLandau, Gumbs, Speth, DiehlEisenrieglerLetter, DiehlEisenrieglerArticle}.  
The latter two references \cite{DiehlEisenrieglerLetter, DiehlEisenrieglerArticle} emphasize a connection to 
polymer physics in the $N=0$ case.  As far as we know,
there is no literature on the large $N$ expansion in the presence of a boundary for $\phi^6$ theory.
It is this gap that the present work attempts to fill.

We are interested in the $O(N)$ scalar field theory with a classically marginal $(\vec \phi^2)^3$ interaction, described by the Lagrangian density
\begin{eqnarray}
\label{phisixL}
{\mathcal L} &=& {\mathcal L}_{\rm bulk} + \delta(z) {\mathcal L}_{\rm bry}\ ,  \\
{\mathcal L}_{\rm bulk} &=&  \frac{N}{2} \left[ (\partial_\mu \vec \phi)^2  + m^2 (\vec \phi^2) + r \, (\vec \phi^2)^2 + \frac{g}{3} (\vec \phi^2)^3  \right] \ , \nonumber \\
{\mathcal L}_{\rm bry} &=& N \left[ h_0 \, \vec \phi \partial_z \vec\phi + h_1 \, \vec \phi^2  +  \frac{h_2}{2} \, (\vec \phi^2)^2 \right] \ , \nonumber
\end{eqnarray}
where $\vec \phi$ is a scalar field with $N$ components and $m$, $r$, $g$, and $h_i$ are couplings.  There is a planar boundary at $z=0$.  We have included all classically relevant and marginal couplings in our Lagrangian density
that preserve the $O(N)$ symmetry.\footnote{%
	Refs.\  \cite{PTandCPreviews,BenhamouMahoux} argue that the $\vec \phi \partial_z \vec \phi$ term
	is in some sense redundant, that having fixed a boundary condition for the field, the coefficient of
	$\vec \phi \partial_z \vec \phi$ becomes 
	scheme dependent and limited in effect to renormalizing the wavefunction of the boundary field $\phi|_{z=0}$.
}$^,$\footnote{%
The particular form of the large $N$ limit we consider here may not be unique. Researchers have speculated about the existence of other large $N$ limits of this theory \cite{Appelquist:1981sf,Osborn:2017ucf}.
}
	
We are especially interested in possible conformal fixed points in three dimensions, and so we tune the relevant mass and interaction couplings, $m$ and $r$, to zero.  For now, we leave the $h_i$ arbitrary as they are useful for controlling the boundary behavior of the fields.
In preparation to do a large $N$ analysis, following \cite{Gudmundsdottir:1985cp}, we rewrite the bulk Lagrangian using two additional Lagrange multiplier fields $\chi$ and $\sigma$:
\begin{equation}
\label{phisixLL}
{\mathcal L}_{\rm bulk} = \frac{N}{2} \left[ (\partial_\mu \vec \phi)^2  + \frac{1}{3} g \chi^3 +  \sigma (\vec \phi^2 - \chi) \right] \ .
\end{equation} 
Integrating over $\sigma$ and then $\chi$ in the path integral restores the Lagrangian density (\ref{phisixL}).  Unlike the usual case of a $\phi^4$ interaction in four dimensions, 
a single Lagrange multiplier field would lead to a nonanalytic interaction term of the form $\sigma^{3/2}$.
We could do something similar for the boundary term ${\mathcal L}_{\rm bry}$ as well, introducing boundary Lagrange multiplier fields $\tilde \chi$ and $\tilde \sigma$, but for now we leave it untouched.

The beta function for this theory without boundary 
was calculated about thirty years ago \cite{Pisarski:1982vz} (see also \cite{Townsend:1976sy,Appelquist:1982vd}):
\begin{equation}
\Lambda \frac{d g}{d \Lambda} = \frac{3 g^2}{2 \pi^2 N} \left( 1 - \frac{g}{192} \right) + O(N^{-2}) \ ,
\end{equation}
indicating that in the large $N$ limit, the beta function approximately vanishes.  We shall take advantage of this fact and treat $g$ as a marginal coupling, to leading order in $1/N$.  The full story is much more interesting and not completely settled.\footnote{%
 See \cite{Omid:2016jve,Fleming:2020qqx} for recent work about this subject.
}  The beta function naively indicates that in the strict large $N$ limit there is a flow from an interacting UV fixed point with $g = 192$ to a free IR fixed point.  In fact, the theory appears to be unstable for $g > 16 \pi^2 \approx 158$
 \cite{Gudmundsdottir:1985cp,Sarbach:1978zz,Bardeen:1983rv}.  We will find some additional evidence for this instability
 from our boundary field theory perspective.
 
We are interested in this particular $\phi^6$ theory because, next to the $\phi^4$ theory mentioned above, it provides  one of the simplest examples of an interacting
boundary conformal field theory (CFT)
in more than two dimensions where explicit calculations can be carried out and the theory examined in detail.  
The boundary CFT aspects of scalar $\phi^4$ theory were well explored in the nineties in two classic papers by McAvity and Osborn \cite{McAvity:1993ue,McAvity:1995zd} using the $\epsilon$ expansion and large $N$ techniques.   (The current work is in fact very heavily influenced in its structure and approach by the latter reference \cite{McAvity:1995zd}.)  More recently, the conformal bootstrap program has provided an additional tool to study these kinds of theories, and there has been a renewed interest in $\phi^4$ theory with a boundary \cite{Liendo:2012hy,Gliozzi:2015qsa,Bissi:2018mcq}. 

Other tractable examples of boundary CFT tend to be more exotic -- they are free in the bulk, or they have supersymmetry, or they are described by a dual gravitational system through the AdS/CFT correspondence.
Regarding theories that are free in the bulk, a close relative of the $\phi^4$ theory with a boundary is a scalar theory that interacts only through the boundary.  See refs.\ \cite{Giombi:2019enr,Prochazka:2019fah} for recent investigations although such a theory provides an important cross check already in \cite{DiehlEisenrieglerArticle}.   Another important class of boundary CFTs that are free in the bulk are graphene like: They have a 4d photon and 3d charged matter 
(see e.g.\ \cite{Herzog:2017xha}).  
The literature about supersymmetric and holographic boundary CFTs we will not attempt to summarize here.

There were two quantities in particular that we sought to compute in looking at this theory, coefficients of the anomaly in the trace of the stress tensor.  
  While the trace of the stress tensor vanishes classically, coupling the theory to a background metric produces anomalous terms in the trace 
  proportional to curvature invariants.  In the absence of a boundary or
  defect, the trace anomaly is present only in even dimensions.  There are however boundary and defect localized contributions to the anomaly in odd dimensions as well.  In the three dimensional case at hand, one finds  
   \cite{Graham:1999pm}
  \begin{equation}
\label{traceanomaly}
\langle T^\mu_{\; \; \mu} \rangle = \frac{\delta(x_\perp)}{4 \pi} \left( a \, R + b \, \hat K_{\mu\nu} \hat K^{\mu\nu} \right) \ ,
\end{equation}
where $\delta (x_\perp)$ is a Dirac delta function with support on the boundary, $\hat K_{\mu\nu}$ is the traceless part of the extrinsic curvature, and $R$ is the Ricci scalar on the boundary.
These coefficients $a$ and $b$ hold promise as a way of classifying and better understanding the properties of 
boundary CFT.  For example, it is known
that $a$ decreases under boundary renormalization group flow \cite{Jensen:2015swa} while $b$
can be computed from the displacement operator two-point function \cite{Herzog:2017kkj}.  
 
 In order to get at these two numbers, we take two different approaches.  The quantity $a$ we obtain by evaluating the partition function of the theory on hyperbolic space.  In section \ref{Sec:O(N)atlargeN}, we use large $N$ methods to compute an effective potential and then finish the computation of $a$ in the discussion in section \ref{sec:disc}.  
 The effective potential also allows us to examine the different possible solutions (or phases) of the theory as a function of the quasi-marginal coupling $g$.  We find an interesting collection of boundary ordered and disordered phases separated by first and second order phase transitions.\footnote{%
 Given the Coleman-Mermin-Wagner Theorem, it may seem surprising that we find boundary ordered phases in our set-up.  From the point of view of the, in general, nonlocal effective two dimensional field theory living on the boundary, this theorem should prohibit surface ordering phase transitions.  Presumably, we find such phases because we are looking in a large $N$ limit.
 }
 
The quantity $b$ we extract from the stress-tensor two-point function.
 In flat space with a boundary at $z=0$, 
 the displacement operator is the boundary limit of the normal-normal component of the stress tensor, 
 $T^{nn}({\bf x}, z)|_{z=0} = D({\bf x})$.  Thus, we can obtain $b$ not only from the two-point function of the displacement operator but also from the boundary limit of the two-point function of the stress tensor.  
 The computation of this two-point function forms the centerpiece of the current work.  We rely heavily on large $N$ techniques and the underlying conformal symmetry of the theory.  Along the way, we also compute the two-point function of the $O(N)$ current operator.  
 
 Given current interest in conformal bootstrap techniques, we analyze also the bulk and boundary conformal block decompositions of our two-point functions.  There are two natural limits of a two-point function in boundary CFT: a coincident limit in which the two insertions get close together and a boundary limit in which at least one of the insertions gets close to the boundary.  In these limits, it is further natural to decompose the operators in an operator product expansion.  In the coincident limit, the decomposition runs over a series of bulk scalar operators.  In the boundary limit, one sums instead over boundary operators.  These decompositions thus give additional information about the operator spectrum and OPE coefficients in the theory.
 
Our work begins in section \ref{Sec:O(N)atlargeN} 
by reviewing how the large $N$ effective Lagrangian is captured by the classical contribution
(\ref{phisixLL}) plus a one loop contribution coming from fluctuations of the $\vec \phi$ field.  We set up some formalism
for calculating Feynman diagrams.  We also analyze how the solution space depends on the coupling $g$.  
We find the rich phase structure summarized in figure \ref{fig:stability}.  
In sections \ref{sec:JJ} and \ref{sec:TT}, we compute the two point functions of the current and stress tensor in the 
Dirichlet boundary case.  
Finally, in section \ref{sec:boundary decomposition}, we decompose these two point functions into series of boundary and bulk conformal blocks, from which we learn something about the spectrum of conformal bulk and boundary primary operators along with their OPE coefficients.
Section \ref{sec:disc} is a discussion of the boundary trace anomaly coefficients that can be deduced from the potential computed in section \ref{Sec:O(N)atlargeN} and the stress tensor two point function computed in section \ref{sec:TT}.
An appendix \ref{App:conformal integral with boundary} contains further details of the stress tensor two-point function calculation.

\section{$O(N)$ model with planar boundary at large $N$ } \label{Sec:O(N)atlargeN}

We begin with a discussion of the boundary conditions. Denoting our coordinate system as $x = (\bm{x},z)$,
we introduce a boundary along the plane $z=0$ so that $\bm{x}$ are tangential to the boundary and $z$ is normal. 
The dominant effect in establishing the boundary conditions is the relevant term $h_1 \vec \phi^2$ in ${\mathcal L}_{\rm bry}$.  The other two operators $\vec \phi \partial_z \vec \phi$ and $(\vec \phi^2)^2$ are marginal.
 In the low energy limit, the effective value of $h_1 / \Lambda$ is $\pm \infty$ or zero.  The case $
h_1 \to \infty$ imposes Dirichlet (or ``ordinary'') conditions on the field $\vec \phi$ while the finely tuned $h_1 = 0$ imposes Neumann (or ``special'').
The case $h_1 \to - \infty$ allows for the so-called extraordinary boundary conditions where $\phi_\alpha \sim z^{-1/2}$. 
Given the Coleman-Mermin-Wagner Theorem, fluctuations should destroy this $\phi_\alpha \sim z^{-1/2}$ ordering
behavior on our two dimensional surface.  We presumably 
see this behavior because we are working in a large $N$ limit where the
fluctuations are suppressed. 

As discussed in \cite{DiehlEisenrieglerLetter, DiehlEisenrieglerArticle}, 
the Neumann case here is more subtle than in $\phi^4$ theory.  At this critical value, the marginal coupling $h_2$ can become important.  These references demonstrated that there is a nonzero beta function for $h_2$, proportional to $g$, in the $3-\epsilon$ expansion.  We do not have much to say about this special case $h_1 = 0$ in the current work, but it would be interesting to examine it more thoroughly in the future.
 
Taking (\ref{phisixLL}) as our starting point, we divide the fields up into background plus fluctuations:
\begin{eqnarray}
\phi_\alpha &=& \delta_{\alpha 1} \frac{\Phi}{z^{1/2}} + \delta \phi_\alpha \ , \\
\sigma &=& \frac{\Sigma}{z^2} + \delta \sigma \ , \\
\chi &=& \frac{\Xi}{z} + \delta \chi \ .
\end{eqnarray}
We are taking advantage of the presence of a boundary at $z=0$ to allow for a coordinate dependence in the background values of the fields.
To find a scale invariant solution, 
we are assuming that at leading order in $N$, the scaling dimensions of $\phi_\alpha$, $\sigma$, and $\chi$ are given by their classical values, and that $\Phi$, $\Sigma$, and $\Xi$ are constants.  We find an effective action for the fluctuations $\delta \phi_\alpha$:
\begin{eqnarray}
\label{fluctL}
\frac{N}{2} \left[ (\partial \delta \phi_\alpha)^2 + \frac{1}{z^2} \Sigma \, \delta \phi_\alpha^2  \right]\ .
\end{eqnarray}
There is a cross term proportional to $\Phi \, \sigma \, \delta \phi_1$ which involves fluctuations only in the direction in which $\phi_\alpha$ is turned on, and thus is down by a power of $1/N$ compared to the expression above; we ignore this cross term.

\subsection{Feynman rules at large $N$}

We begin with an analysis of the Lagrangian density (\ref{fluctL}) which describes the behavior of a free scalar field with a position dependent mass.  
The $O(N)$ symmetry 
 restricts the form of two-point functions to be $\langle \delta \phi_\alpha (x) \delta \phi_\beta (x') \rangle = \delta_{\alpha\beta} G_\phi (x, x')$, and then $G_\phi$ can be determined by
\begin{align}
    \left[  \Box - \frac{\mu^2 - \frac{1}{4}}{z^2}\right]G_\phi (x, x') = \delta(x-x') \ , \; \; \; \label{equation for two-pt. functon} 
    \Sigma \equiv \mu^2 -\frac{1}{4} \ .
\end{align}
The Lagrangian density (\ref{fluctL}), including the position dependent mass, preserves a $SO(4,1)$ symmetry associated with a Euclidean boundary conformal field theory in three dimensions.  As it is not more difficult, let us work in general dimension.  The symmetry implies 
that $G_\phi$ must take the form \cite{McAvity:1995zd},  
\begin{align} \label{general form of two-pt of phi}
    G_\phi (x,x') = \frac{F(v)}{|x-x'|^{d-2}} \, ,
\end{align}
where $v$ is the conformal cross ratio given by
\begin{align}
    v^2 \equiv \frac{\left( {\bm x} - {\bm x'}\right)^2 + (z-z')^2}{\left( {\bm x} - {\bm x'}\right)^2 + (z+z')^2} \, ,
\end{align}
and we used the fact that at leading order the bulk scaling dimension of 
$\delta \phi_\alpha$ is given by $\Delta_\phi = d/2 - 1$. Given \eqref{equation for two-pt. functon} and \eqref{general form of two-pt of phi}, we see that $F(v)$ satisfies the differential equation, 
\begin{align} \label{equation for bdy block}
    (1-v^2)^2 v F''(v) - (d-3)(1-v^2)^2 F'(v) - \left( 4\mu^2 -1 \right) v F(v) = 0 \, .
\end{align}

To have a well defined problem, we need to fix the boundary conditions in the coincident $v=0$ and boundary $v=1$ limits.
In the coincident limit, we expect to recover the usual two-point function for a massless free field, 
\begin{align}
    G_\phi (x, x')  \sim \frac{\kappa}{|x-x'|^{d-2}} \, , \; \; \; \kappa \equiv \frac{1}{N (d-2) \Omega_{d-1}}  \ ,
\end{align}
where the value of $\kappa$ follows from the normalization of the kinetic term for $\delta \phi_\alpha$.  
That the Lagrangian has an over-all factor of $N$ means the propagators must all scale with $1/N$.  
Note also $\Omega_d$ is the volume of a unit $d$ dimensional sphere.  

In the boundary limit where $v \rightarrow 1$, there are two possible behaviors $F(v) \sim(1-v^2)^{\frac{1}{2} \pm \mu}$.  We keep the $+ \mu$ behavior and set the other scaling behavior to zero; a linear combination would force us to introduce a scale and break the conformal symmetry. 
Note the choice $\mu = \frac{1}{2}$ is the usual Dirichlet boundary condition while $\mu = - \frac{1}{2}$ is Neumann.
With these boundary conditions,  the unique solution of \eqref{equation for bdy block} is 
\begin{align} \label{bdy block}
    F(v) = \kappa \frac{\Gamma(\frac{1}{2}+\mu) \Gamma(\frac{d-1}{2}+\mu ) } { \Gamma(\frac{d}{2}-1) \Gamma(1 + 2 \mu)} \xi^{-\frac{1}{2}-\mu} {}_2 F_1 \left(\frac{1}{2} + \mu, \, \frac{d-1}{2} + \mu, \, 1 + 2\mu, \, - \frac{1}{\xi}  \right) \, ,
\end{align}
where $\xi$ is a different expression of the cross ratio related to $v$ as 
\begin{align}
    v^2 = \frac{\xi}{\xi + 1} \, .
\end{align}
We can of course recover the other boundary condition at $z=0$ by changing the sign $\mu \to - \mu$.

Finally we comment on the propagators of auxiliary fields $\sigma$ and $\chi$.
The equation of motion for $\sigma$ states that $\vec \phi^2 - \chi = 0$.  By the Schwinger-Dyson equations, any correlation function involving this equation of motion should vanish up to contact terms.  In particular, we have
\begin{align}
 \frac{N}{2} \langle \sigma(x) (\vec \phi^2(x') - \chi(x'))  \rangle = \delta(x-x') \ .
\end{align}
We expect in the large $N$ limit that the $\langle \sigma(x) \vec \phi^2(x') \rangle$ piece of the expression dominates as there
are $N$ identical components of $\vec \phi$.  Furthermore, we can re-express this three point function in terms of the corresponding propagators and the three point vertex $\frac{N}{2} \sigma \vec \phi^2$ in the effective Lagrangian.
\begin{align} \label{phi^2 and sigma 3-pt.}
\langle \phi_\alpha (x_1) \phi_\beta(x_2) \sigma(x_3) \rangle = -\delta_{\alpha\beta} N \int_{{\mathbb R}_+^d} \d^d r \, G_\phi(x_1, r) G_\phi(x_2, r) G_\sigma(r,x_3) \ .
\end{align}
In particular, we learn that
\begin{align} \label{basic eq for sigma}
 \int_{{\mathbb R}_+^d} \d^d x'' G_\phi^2 (x,x'') G_\sigma (x'',x') = -\frac{2}{N^3} \delta^d(x-x') \ .
\end{align}
Given that $G_\phi$ is $O(1/N)$, we conclude that $G_\sigma$ is also $O(1/N)$.  
We don't need the explict form of $G_\sigma$, but will make heavy use of \eqref{basic eq for sigma} later. 

\begin{figure}
\begin{center}
a)\includegraphics[width=1.5in]{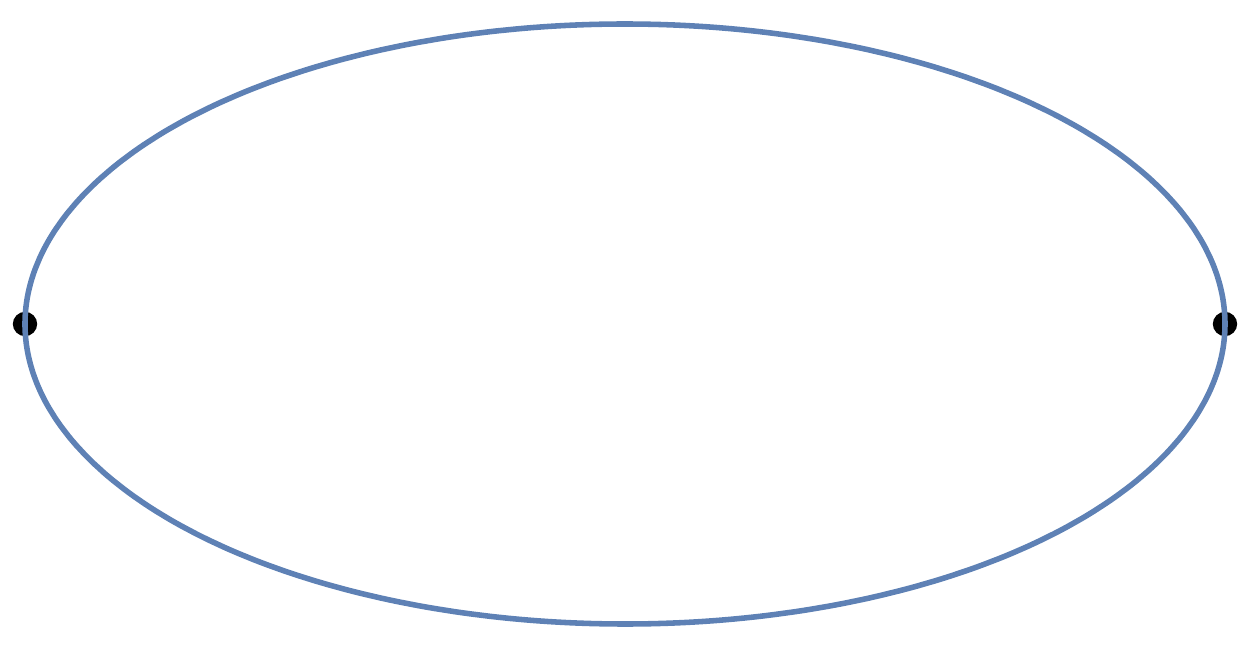}
b)\includegraphics[width=1.5in]{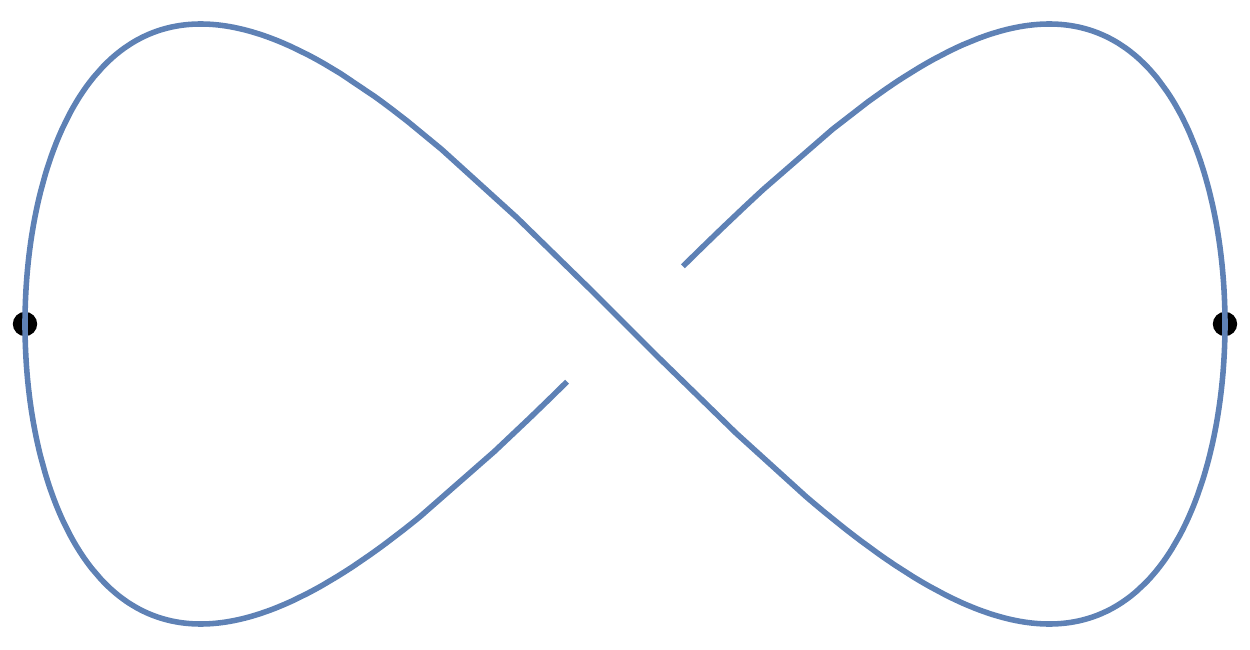}
c)\includegraphics[width=1.5in]{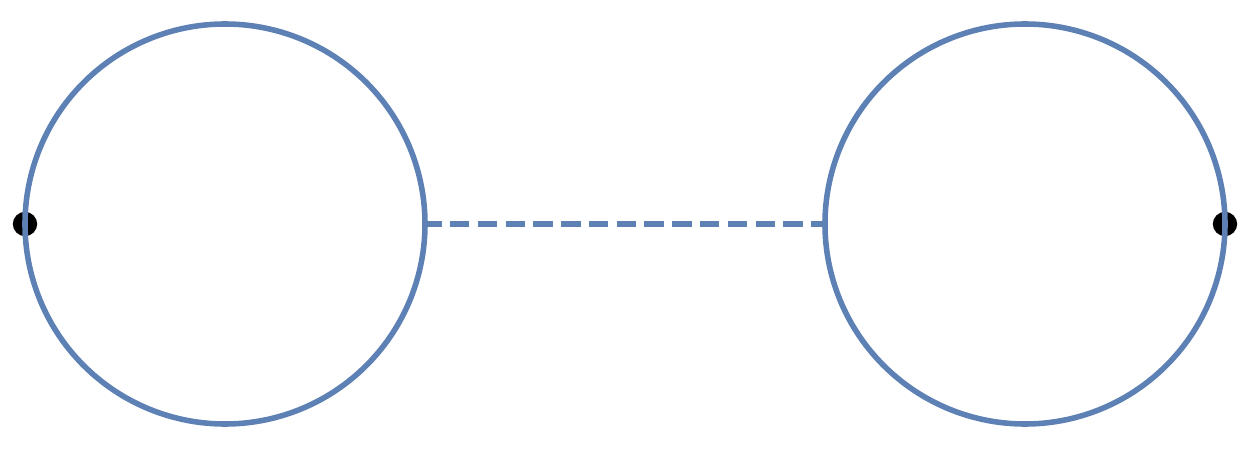}
\end{center}
\caption{
The Feynman diagrams needed for computing the $\langle J^{\alpha\beta}_\mu (x) J^{\gamma \delta}_\nu(x')  \rangle$ and $\langle T^{\mu\nu}(x) T^{\lambda \rho}(x') \rangle$ correlation functions at leading order in $N$.  Diagrams (a) and (b) contribute to $\langle J^{\alpha\beta}_\mu (x) J^{\gamma \delta}_\nu(x') \rangle$ at order $N^{0}$ while (c) vanishes because of the antisymmetrization over the $O(N)$ indices.  For  $\langle T^{\mu\nu}(x) T^{\lambda \rho}(x') \rangle$, all three diagrams contribute an amount proportional to $N$.  The solid lines are $\phi_\alpha$ propagators while the dashed line is a $\sigma$ propagator.
}
\label{fig:leadingdiagrams}
\end{figure}

The diagrams in figure
\ref{fig:leadingdiagrams} give the leading contributions to the current and stress tensor correlation functions of interest.  We use rules where every propagator comes with a factor of $1/N$, every vertex 
and every loop with a factor of $N$.  The black dots correspond to the inserted operators and 
may influence the $N$ counting. 

In the next subsection, we will use the $G_\phi$ propagator to compute a one-loop effective potential while in sections \ref{sec:JJ} and \ref{sec:TT}, we will use these Feynman rules to study the $\langle J^{\alpha \beta}_\mu (x) J^{\gamma \delta}_\nu(x') \rangle$  and $\langle T^{\mu\nu}(x) T^{\sigma \rho}(x') \rangle$ correlation functions at leading order in $N$.

\subsection{Effective potential}

The quantum fluctuations from the $\delta \phi_\alpha$ fields modify the original Lagrangian by a one loop effect:
\begin{equation}
\label{Lcorrection}
 {\mathcal L}  \to {\mathcal L} + \frac{N}{2} \tr \log \left( - \Box + \frac{\Sigma}{z^2} \right) \ .
\end{equation}
The trace log factor is the integral of the one-point function of the operator $\langle \delta \phi_\alpha^2 \rangle$.  
We can construct this one-point function from the regulated coincident limit of the Green's function $G_\phi(x,x')$.
By a hypergeometric identity, the result (\ref{bdy block}) can be written as  
\begin{align}
F(v)  = & \kappa (1-v^2)^{\frac{1}{2}-\mu} {}_2 F_1 \left( \frac{1}{2}- \mu, \frac{3-d}{2} - \mu, 2 - \frac{d}{2}, v^2 \right) \nonumber \\
& + c v^2  (1-v^2)^{\frac{1}{2}-\mu} {}_2 F_1 \left( \frac{1}{2} - \mu, \frac{d-1}{2} - \mu, \frac{d}{2}, v^2 \right) \ ,
\end{align}
where
\[
c = \kappa \frac{\Gamma \left( 1 - \frac{d}{2} \right) \Gamma \left( \frac{d-1}{2} + \mu \right)}
{\Gamma \left(\frac{d}{2}-1 \right) \Gamma \left( \frac{3-d}{2} + \mu \right) } \ .
\]
The first hypergeometric function has singularities that must be removed in the coincident limit $v \to 0$.  The one-point function is then fixed essentially by the constant $c$:
\begin{align}
\langle \delta \phi_\alpha^2 (x) \rangle = \frac{\kappa}{N} 
\frac{\Gamma \left( 1 - \frac{d}{2} \right) \Gamma \left( \frac{d-1}{2} + \mu \right)}
{2^{d-2} \Gamma \left(\frac{d}{2}-1 \right) \Gamma \left( \frac{3-d}{2} + \mu \right) } \frac{1}{z^{d-2}}  \, , 
\end{align}
summation on $\alpha$ not implied.

Integrating this one-point function over $\mu$ gives the difference in effective potential between theories with different values of $\mu$:
\begin{align}
\frac{ N \kappa }{z^{d}} \int_0^\mu\frac{\Gamma \left(1 - \frac{d}{2} \right) \Gamma \left( \frac{d-1}{2} + x \right)}{ 2^{d-2} \Gamma \left(\frac{d}{2}-1 \right) \Gamma \left(\frac{3-d}{2} + x \right)} x \, \d x\ .
\label{tracelog}
\end{align}
We followed \cite{Herzog:2019bom} in this derivation but see also 
\cite{McAvity:1995zd,Carmi:2018qzm}.\footnote{%
For (\ref{tracelog}) to be consistent with scale invariance, we must either be in $d=3$ dimensions or the integral must vanish.
We are in $d=3$, but it is useful to compare with the general $d$ results of other authors.
The large $N$ results of Bray and Moore \cite{Bray:1977tk} and later McAvity and Osborn \cite{McAvity:1995zd} correspond to setting the integrand to zero which happens when $\mu = \frac{d-3}{2}, \frac{d-5}{2}, \frac{d-7}{2}$, etc.  The first two cases are the ``ordinary'' (Dirichlet) and ``special'' (Neumann) phase transitions close to $d=4$.  
In general, the scaling $\mu$ means there is an operator on the boundary with scaling dimension $\mu + \frac{d-1}{2}$.
The condition the integrand vanishes gives the series of dimensions $d-2$, $d-3$, $d-4$, etc.
The unitarity bound cuts off this series at $d-3$ in $d=4$ and at $d-2$ in $d=3$.  
}
Note we are using $\mu = 0$ as a reference value around which to compute the change in the potential.

In the context of the relevant $h_1 \, \vec \phi^2 \, \delta(z)$ deformation that sets the boundary condition, we have three cases in which to consider values of $\mu$.  In the Dirichlet case $h_1>0$, provided $\mu > -1/2$, the boundary condition $\phi_\alpha = 0$ remains untouched.  In the extraordinary case $h_1<0$, there is no constraint on $\mu$ as $\phi_\alpha$ is already infinite on the boundary.  Finally, there is the finely tuned ``Neumann'' case $h_1=0$, 
for which further analysis is needed to sort out the role of the $h_0$ and $h_2$ boundary couplings, analysis which
we leave for the future.

For us, in $d=3$, the expression (\ref{tracelog}) reduces to $-\frac{N \mu^3}{12 \pi z^3}$.  The equations of motion give the following conditions on $\Phi$, $\Sigma$, and $\Xi$:
\begin{eqnarray}
\Phi (3 - 4 \Sigma) &=& 0 \ , \nonumber \\
\label{myeom}
\pm \sqrt{1+4 \Sigma} -8 \pi(\Phi^2 - \Xi) &=&  0 \ , 
\label{criteqs}
\\
\Xi^2 g - \Sigma &=& 0 \ , \nonumber
\end{eqnarray}
where the $\pm$ in the second line corresponds to a choice of sign for $\mu$.
The boundary ordered and disordered solutions to these three equations are summarized in figure \ref{fig:solutions}.  We will discuss how to compute the potential $V$ in this figure shortly.

\begin{figure}
\begin{align}
\begin{array}{|c|c|c|c|c|}
\hline
& \multicolumn{2}{|c|}{\mu>0} & \multicolumn{2}{c|}{\mu<0} \\
\hline
\Phi & 0 & \frac{1}{2 \sqrt{2 \pi}} \sqrt{1 \pm \frac{2 \pi \sqrt{3}}{\sqrt{g}}}&  0 &  
\frac{1}{2 \sqrt{2 \pi}} \sqrt{-1 + \frac{2 \pi \sqrt{3}}{\sqrt{g}}}\\
\hline
\Sigma & \frac{g}{4 (16 \pi^2 - g)} & \frac{3}{4} &  \frac{g}{4 (16 \pi^2 - g)} & \frac{3}{4}  
\\
\hline
\Xi & - \frac{1}{2 \sqrt{16 \pi^2 - g}} & \pm \sqrt{\frac{3}{4g}} &  \frac{1}{2 \sqrt{16 \pi^2 - g}} & \sqrt{\frac{3}{4g}}
\\
\hline
\frac{2}{N} V & 
- \frac{1}{12 \sqrt{16 \pi^2 - g}}
&
- \frac{1}{6 \pi} \pm \sqrt{\frac{3}{16 g}} 
&
 \frac{1}{12 \sqrt{16 \pi^2 - g}}
 &
 \frac{1}{6 \pi} - \sqrt{\frac{3}{16 g}}
 \\
 \hline
 \end{array} \nonumber
\end{align}
\caption{The various solutions to the equations (\ref{criteqs}).  The potential $V$ is calculated from (\ref{Veff}).}
\label{fig:solutions}
\end{figure}

There are boundary ordered phases with $\phi_\alpha \neq 0$.  
There are two such solutions with $\mu>0$.  The solution associated with negative $\Xi$ exists only for 
$g > 12 \pi^2$ and corresponds to a local maximum of the effective potential, as we will see shortly.  
For negative $\mu$, there is only a single ordered solution, and it 
exists only for $g < 12 \pi^2$.  
Note $\Sigma = 3/4$ corresponds to $\mu = \pm 1$.  
The value $g = 12 \pi^2$ is special for another reason, for here two of the three boundary ordered phases become disordered,
with $\phi_\alpha = 0$.  

There are a pair of disordered solutions with $\phi_\alpha = 0$ for more general values of $g$, 
one for each sign choice of $\mu$.
Note $\mu^2 = (4 - g/4 \pi^2)^{-1}$ for these solutions.  The dependence of $\Xi$ on $g$ in the 
disordered phase, in particular that $\Xi$ becomes imaginary for $g > 16 \pi^2$,
 suggests the theory becomes sick for $g > 16 \pi^2$, 
consistent with the results \cite{Gudmundsdottir:1985cp,Sarbach:1978zz,Bardeen:1983rv} 
in absence of a boundary.

For comparison, we  can work in a Weyl equivalent frame where the fields take constant rather than $z$ dependent values.  That frame is three dimensional hyperbolic space $H_3$ with radius of curvature $L$.  We must remember to include the conformal coupling 
of $\phi_\alpha$ to the curvature ${\mathcal L } \to {\mathcal L} +\frac{N}{2}  \frac{d-2}{4(d-1)} R \phi_\alpha^2$ where $R = - \frac{d(d-1)}{L^2}$ for H$_d$ where $L$ is the radius of curvature.  In our particular case, we are adding a mass term $-\frac{N}{2} \frac{3}{4 L^2} \phi_\alpha^2$ in $H_3$.    
We find the following effective potential for the fields
\begin{equation}
\label{Veff}
V = \frac{N}{2} \left[ \frac{1}{3} g \Xi^3 + \Sigma (\Phi^2 - \Xi) - \frac{3}{4} \Phi^2 \mp \frac{(\Sigma + \frac{1}{4})^{3/2}}{6 \pi} \right] \, ,
\end{equation}
which gives rise to the same conditions (\ref{myeom}).  The choice in sign refers to the choice of sign of $\mu$.  From this hyperbolic viewpoint, we should keep the mass of the scalar field above the Breitenlohner-Freedman  bound, $\Sigma - \frac{3}{4} > -1$.  In the disordered phase, for $g$ in the allowed range $-\infty < g < 16 \pi^2$, $\Sigma$ satisfies the bound, while for $g > 16 \pi^2$, the fluctuations in the scalar field will have a mass below the BF bound, and the theory should be unstable.

\begin{figure}
\begin{center}
\includegraphics[width = 6in]{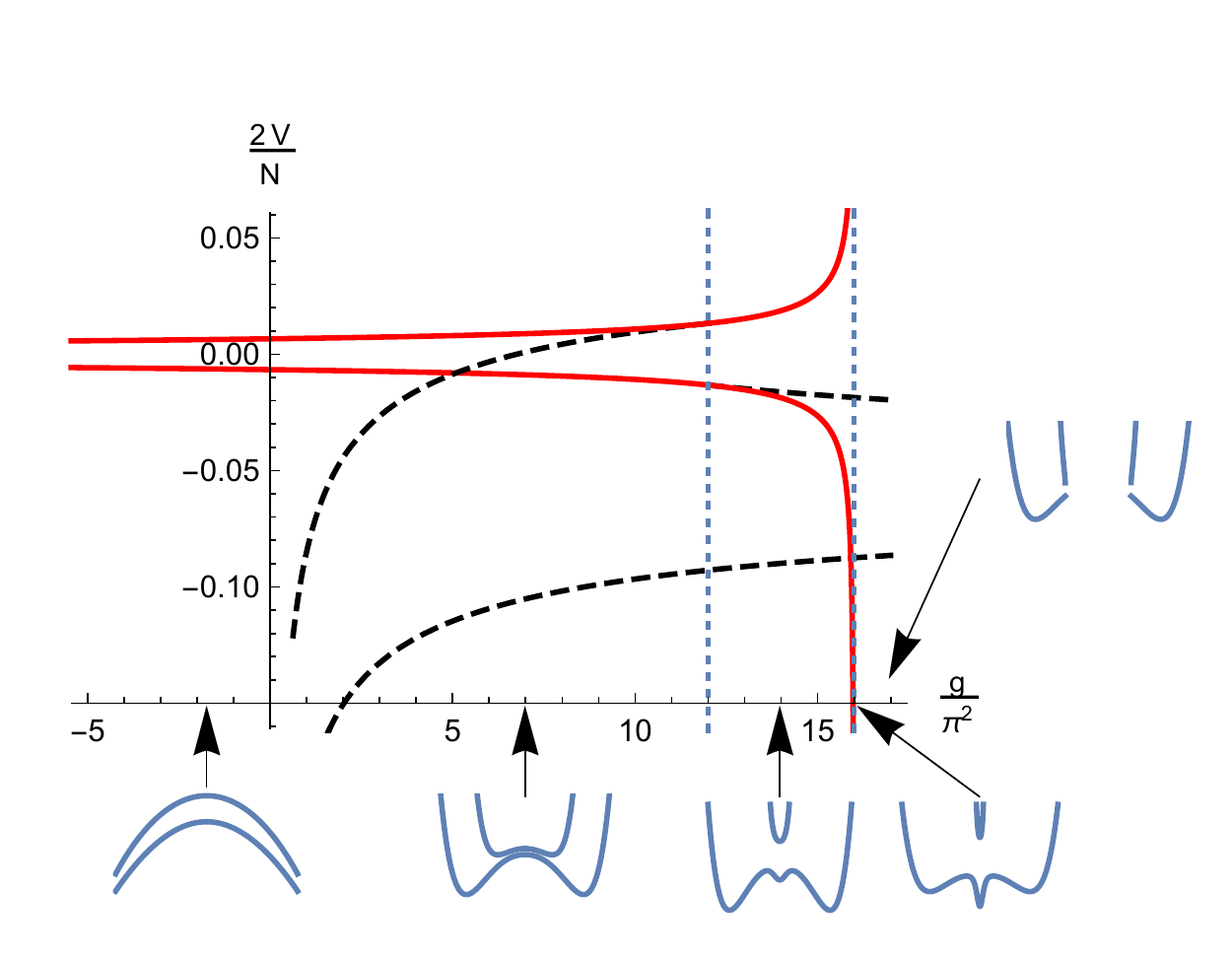}
\end{center}
\caption{Potential vs.\ coupling. 
The solid curves are disordered ($\phi_\alpha = 0$) and the dashed curves are boundary ordered ($\phi_\alpha \neq 0$).  
The disordered phases cease to exist for $g > 16 \pi^2$ while the ordered phases require $g > 0$.  
The disordered phases can join with the ordered phases at $g = 12 \pi^2$.  
Dotted vertical lines are placed at
$g = 12 \pi^2$ and $g = 16 \pi^2$ as a guide to the eye.  
The inset plots show the qualitative shape of the potential as a function of $\Phi$ in the different regions of the larger plot.
There are two different branches of $V(\Phi)$: the upper branch corresponds to $\mu<0$ and the lower branch to $\mu>0$.
 \label{fig:stability}}
\end{figure}

To understand relative stability of the different phases, we can study the potential $V$ (see figure \ref{fig:stability}).  
The analysis has some familiar Landau-Ginzburg features, but 
is complicated by the dependence of the phases on boundary conditions.  
One can form an effective potential $V(\Phi)$ of a single variable by first extremizing $V(\Phi, \Sigma, \Xi)$ with respect to $\Sigma$ and $\Xi$.  We find that for $g <0$, the potential has a single maximum, albeit with a curvature below the BF bound.  For $0 < g < 12 \pi^2$, the potential has a classical Mexican hat shape, with minima corresponding to the ordered phase and a maximum corresponding to the disordered phase.  Then for $12 \pi^2 < g < 16 \pi^2$, there is a qualitative difference between the $\mu>0$ and $\mu<0$ cases.  For $\mu>1$, the maximum at $\Phi = 0$ develops a dimple that grows deeper and eventually overtakes the minima associated with the ordered phase.  In constrast, for $\mu<-1$, the disordered and ordered phases coalesce into a single minimum associated with a stable disordered phase.
Given that $\mu<-1$ leads to a surface primary below the unitarity bound, we could discard this portion of the $\mu<0$ disordered phase based on unitarity.
For $g > 16 \pi^2$ and either choice of sign for $\mu$, the effective potential $V(\Phi)$ is not defined for $\Phi$ close to the origin although there are still critical points associated with the disordered phases.  

Recall that we impose a boundary condition on the field $\vec \phi$ by adding the relevant boundary deformation 
$h_1 \, \vec \phi^2 \, \delta(z)$.  For $h_1>0$, $\vec \phi$ must vanish on the boundary.  To be consistent with this Dirichlet condition, the critical exponent for the fluctuation $\delta \phi_\alpha$ must satisfy $\mu > -1/2$.  
The only phases that are consistent with these restrictions are the 
lower solid (red) curve in figure \ref{fig:stability} and the portion of the upper solid (red) curve satisfying $g < 0$.  As the lower curve has lower potential $V$, it should represent the stable phase. 

We next consider the choice $h_1<0$, for which $\vec \phi$ can blow up at the boundary -- extraordinary boundary conditions.  In this case, as $\vec \phi$ is already infinite, there is no restriction on $\mu$ of the fluctuation field $\delta \phi_\alpha$.  
All of the curves in figure \ref{fig:stability} are allowed.  Based on energetic considerations, the lower dashed (black) curve, corresponding to a boundary ordered phase, is preferred in the range $0 < g < \frac{3}{2} ( 7 + \sqrt{13}) \pi^2 \approx 15.9 \pi^2$.  At the upper end of the range, there is a first order phase transition to a boundary disordered phase.  For
$ \frac{3}{2} ( 7 + \sqrt{13}) \pi^2 < g < 16 \pi^2$, the boundary disordered phase is preferred.  
In the regime $g < 0$, there are only boundary disordered phases, while in the regime $g > 16 \pi^2$ there are only boundary ordered phases.  (Given the Coleman-Mermin-Wagner Theorem, 
we should of course keep in mind that we are likely only seeing 
boundary ordered phases because of the large $N$ limit.)

The last case is ``Neumann'' boundary conditions $h_1=0$.  In reality, at this point the marginal couplings $h_0$ and $h_2$ become important, and the system needs a more thorough examination.  For this reason, we put ``Neumann'' in parentheses because the actual boundary conditions will be determined by $h_0$ and $h_2$.  We leave a more thorough examination of this case to the future.

We note before moving on that 
it is not clear to us that  the theory makes sense outside the range
$0 \leq g < 16 \pi^2$.  The potential is unbounded for $g < 0$ and missing pieces for $g \geq 16 \pi^2$.

\section{Two-point function of the conserved current at large $N$}
\label{sec:JJ}

We compute the current two-point function
$\langle J_\mu^{\alpha \beta} (x) J_\nu^{\gamma \delta} (x') \rangle$ in the Dirichlet boundary 
case $\Phi = 0$.
Since the model has $O(N)$ global symmetry, we have the associated conserved current $J_\mu^{\alpha\beta} = - J_\mu^{\beta \alpha} $.
From Noether's theorem, this current is 
\begin{equation}
J^{\alpha \beta}_\mu =  N ( \phi^\alpha \partial_\mu \phi^\beta - \phi^\beta \partial_\mu \phi^\alpha ) \ . 
\end{equation}
The over all factor of $N$ comes from the normalization of our Lagrangian.
At large $N$, the leading contribution to the two-point function comes from Wick's theorem, i.e.\ figures \ref{fig:leadingdiagrams}a and \ref{fig:leadingdiagrams}b:
\begin{align}
  \langle J_\mu^{\alpha \beta} (x) J_\nu^{\gamma \delta} (x') \rangle &= (\delta^{\alpha \gamma}\delta^{\beta \delta} - \delta^{\alpha \delta} \delta^{\beta \gamma} ) G_{J, \mu \nu} (x, x')\nonumber  \, , \\
  G_{J, \mu \nu} (x, x') &=
    N^2 \left[ G_{\phi}(x,x') \partial_\mu \partial_\nu ' G_\phi (x,x') - \left(\partial_\mu G_\phi (x,x') \right) \left( \partial_\nu ' G_\phi (x,x' )\right)  \right]
   \, . \label{two-pt of current}
\end{align}
On the other hand, by conformal symmetry  \cite{McAvity:1995zd} 
we know the two-point function of the conserved current has the following form,
\begin{align}
  G_{J, \mu \nu} = \frac{1}{(s^2)^{d-1}} \left(I_{\mu \nu} C(v) + X_\mu X_\nu ' D(v) \right) \, .
\end{align}
We have introduced several structures here, first among them the difference vector $s_\mu \equiv x_\mu - x'_\mu$. 
We also have the bitensor
\begin{align}
I_{\mu\nu} \equiv \delta_{\mu\nu} - \frac{2 s_\mu s_\nu}{s^2} \ ,
\end{align}
and the vectors
\begin{align}
X_\mu \equiv v \left(\frac{2z}{s^2} s_\mu - n_\mu \right) \ , \; \;
X'_\mu \equiv v \left(- \frac{2z'}{s^2} s_\mu - n_\mu \right) \ ,
\end{align}
where $n_\mu$ is a unit normal to the boundary.
Comparing \eqref{general form of two-pt of phi} and \eqref{two-pt of current}, we deduce
\begin{align}
\label{Ceq}
  C(v) &=
  (d-2) F(v)^{2}-\left(1-v^{2}\right) v F(v) F^{\prime}(v) \ ,
   \\
   \label{Deq}
  D(v) & = 
  v F(v) \frac{\mathrm{d}}{\mathrm{d} v}\left(\left(1-v^{2}\right) v F^{\prime}(v)\right)-v^{2}\left(1-v^{2}\right) F^{\prime}(v)^{2}
  \ . 
\end{align}

The conservation Ward identity for the current-current two-point function implies that
\begin{equation}
\label{JJWard}
v (C'(v) + D'(v))  = (d-1) D(v) \ .
\end{equation}
One can check that $C(v)$ and $D(v)$ satisfy this relation, for any $d$.  This check is in contrast
to what happens for the stress-tensor two point function, where it is important to include also a diagram
that involves $\sigma$ exchange to recover the conservation Ward identity.

\begin{figure}
  \centering
  \includegraphics[width = 12cm]{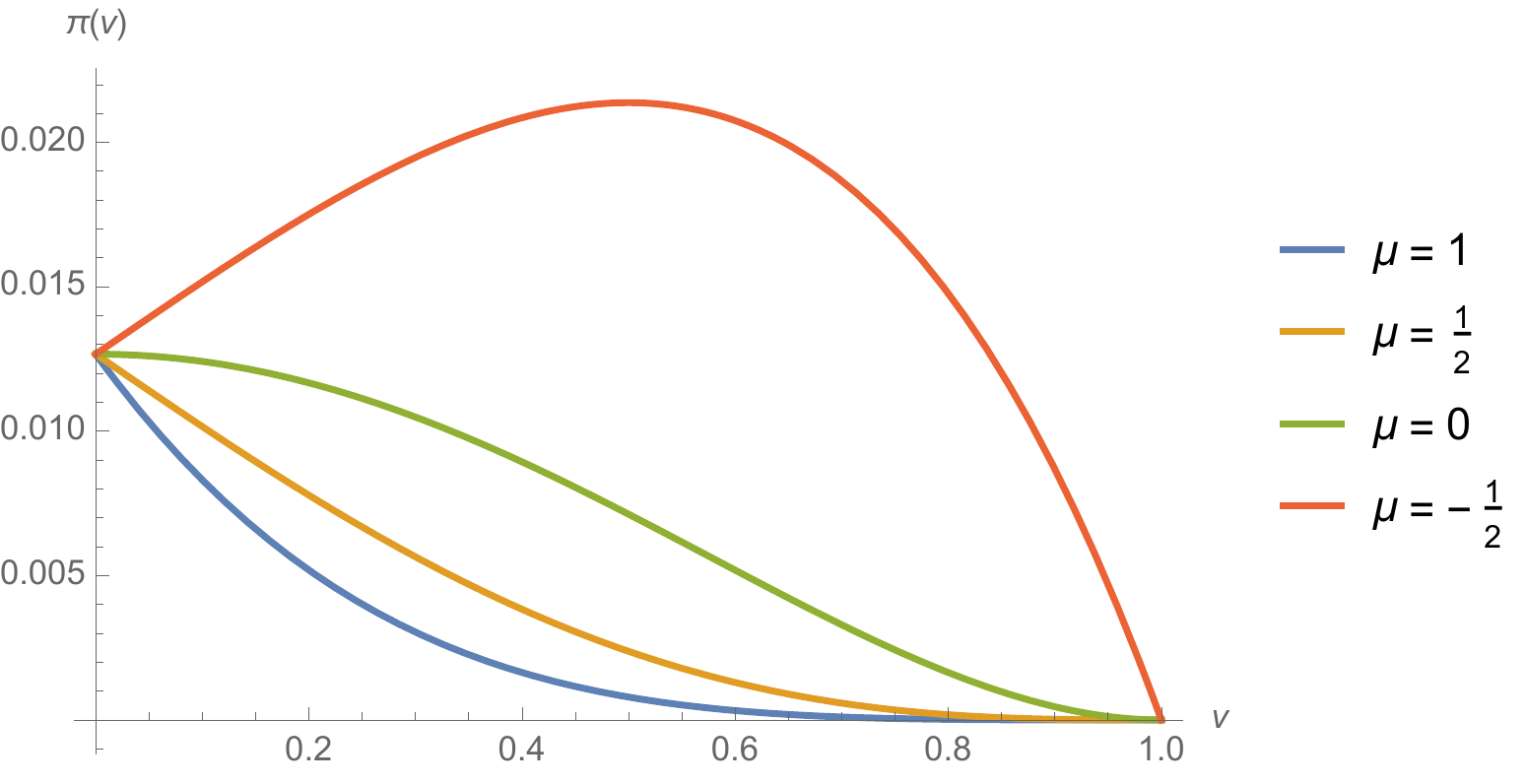}
  \caption{This plot shows when $\mu$ is non-negative, $\pi(v)$ monotonically decreases while $\pi(v)$ has an extremum for negative $\mu$. }
  \label{fig:pis}
\end{figure}

In $d=3$, using \eqref{bdy block} we end up with 
\begin{align}
  C(v) & = C_J (1-v)^{2 \mu +1} (v+1)^{1-2 \mu } \left(v^2+2 \mu  v+1\right) \, , \\
  \pi (v) &\equiv  C(v) + D(v) = C_J (1-v)^{2(1 + \mu)} (1 + v)^{2(1 - \mu)} \, , \label{pi}
\end{align}
where 
$C_J = \frac{2}{(d-2)\Omega_{d-1}^2}$ and 
where $\pi(v)$ is defined such that
\begin{align}
  G_{J,nn} ({\bm 0}, z,{\bm 0}, z') = \frac{\pi (v)}{(z-z')^{2(d-1)}} \, .
\end{align}
We plot $\pi(v)$ for several $\mu$'s in figure \ref{fig:pis}. We find that for non-negative $\mu$, $\pi(v)$ is a monotonically decreasing function of $v$. In contrast, when $0>\mu > -1$, $\pi(v)$ first increases and then decreases once $v$ is large enough. (For $\mu\leq-1$,  $\pi(v)$ monotonically increases, but the unitarity bound for boundary scalar operators implies $\mu \geq -1$.) 

\section{Two-point function of stress tensor at large $N$}
\label{sec:TT}

Here we compute the stress tensor two point function $\langle T^{\mu\nu}(x) T^{\sigma \rho}(x') \rangle$ in the
Dirichlet boundary case $\Phi = 0$.
The stress tensor for the conformally coupled scalar $\vec \phi$ in the presence of the position dependent coupling due to the $\sigma$ field is
\begin{align}
\label{Tmunu}
\frac{1}{N} T_{\mu\nu} =& (\partial_\mu \vec \phi)\cdot (\partial_\nu \vec \phi) - \frac{\delta_{\mu\nu}}{2} \left( (\partial \vec \phi)^2 + \left( \mu^2 - \frac{1}{4} \right) \frac{\vec \phi^2}{z^2} \right) - \frac{d-2}{4(d-1)} \left( \partial_\mu \partial_\nu - \delta_{\mu\nu} \partial^2 \right) \vec \phi^2 \ \nonumber \\
=& - \vec \phi \cdot {\mathcal D}_{\mu\nu} \vec \phi + 
\frac{d}{4(d-1)}{\mathcal D}_{\mu\nu} \vec \phi^2 -
\frac{\delta_{\mu\nu}}{d} \left(\mu^2 - \frac{1}{4} \right) \frac{\vec \phi^2}{z^2} \ ,
\end{align}
where in the last line we used the equation of motion and introduced ${\mathcal D}_{\mu\nu} \equiv \partial_\mu \partial_\nu - \frac{1}{d} \delta_{\mu\nu} \partial^2$.  The overall factor of $N$ comes from the normalization of the Lagrangian.

Using the usual Feynman rules adapted to this large $N$ boundary situation, 
we divide up the calculation of the stress-tensor two point function into a free part and an interaction part:
\begin{align}
\langle T^{\mu\nu}(x) T^{\sigma \rho}(x') \rangle = \langle T^{\mu\nu}(x) T^{\sigma \rho}(x') \rangle_{\rm free} + \langle T^{\mu\nu}(x) T^{\sigma \rho}(x') \rangle_{\rm int} \ .
\end{align}

For the free part, we use the stress tensor (\ref{Tmunu}) and Wick's Theorem, albeit with the propagator $G_\phi(x,x')$ involving a nonzero $\mu$.  The two different ways of contracting the $\phi$ fields give the $t$ and $u$ channel diagrams in figure \ref{fig:leadingdiagrams}.  We can further decompose the free contribution into a trace free part
\begin{align}
 \label{free part of TT}
\frac{1}{N^3} \langle T^{\mu\nu}(x) T^{\sigma \rho}(x') \rangle'_{\rm free} =& \, 
G_\phi {\mathcal D}_{\mu\nu} {\mathcal D}'_{\sigma\rho} G_\phi + ({\mathcal D}_{\mu\nu} G_\phi) {\mathcal D}'_{\sigma \rho} G_\phi \nonumber \nonumber \\
&
 - \frac{d}{2(d-1)} \left({\mathcal D}_{\mu\nu} (G_\phi {\mathcal D}'_{\sigma \rho} G_\phi) 
+ {\mathcal D}'_{\sigma \rho} (G_\phi {\mathcal D}_{\mu\nu} G_\phi ) \right) \nonumber \\
& + \frac{d^2}{8(d-1)^2} {\mathcal D}_{\mu\nu} {\mathcal D}'_{\sigma \rho} G_\phi^2 \ ,
\end{align}
and a remainder
\begin{align}
 \langle & T^{\mu\nu}(x) T^{\sigma \rho}(x') \rangle_{\rm free} -   \langle T^{\mu\nu}(x) T^{\sigma \rho}(x') \rangle'_{\rm free} =  \\
 &-  \frac{2}{d} \left(\mu^2 - \frac{1}{4} \right) \left( \frac{\delta_{\mu\nu} \hat t_{\sigma \rho}(x',x)}{z^2} 
 + \frac{\delta_{\sigma \rho} \hat t_{\mu\nu} (x,x')}{z'^2} \right) + \frac{2N^3}{d^2} \delta_{\mu\nu} \delta_{\sigma \rho} \frac{\left(\mu^2 - \frac{1}{4}\right)^2}{(zz')^2} G_\phi^2(x,x') \ , \nonumber
\end{align}
where we have defined
\begin{align}
\label{tmunudef}
\frac{1}{N^3} \hat t_{\mu\nu} (x,x') \equiv -G_\phi(x,x') {\mathcal D}'_{\mu\nu} G_\phi(x,x') + \frac{d}{4(d-1)} {\mathcal D}_{\mu\nu} (G_\phi(x,x'))^2 \ .
\end{align}

The interaction contribution to the stress-tensor is dominated at leading order in $N$ by exchange of a $\sigma$ field:
\begin{align}
\langle T^{\mu\nu}(x) T^{\sigma \rho}(x') \rangle_{\rm int}  =  \int_{\mathbb R^d_+} \d^d r \int_{\mathbb R^d_+} \d^d r' \, 
t_{\mu\nu}(x,r) t_{\sigma \rho}(x',r') G_\sigma(r,r') \, ,
\end{align}
where the unhatted $t_{\mu\nu}(x,x')$ has a trace part,
\begin{align}
t_{\mu\nu}(x,x') = \hat t_{\mu\nu}(x,x') - \frac{\delta_{\mu\nu}}{d} \frac{\mu^2 - \frac{1}{4}}{z^2} N^3 G_\phi(x,x')^2 \ .
\end{align}
Because of the identity (\ref{basic eq for sigma}), the trace parts of the free contribution and the interaction contribution cancel out and one is left with
\begin{align}
\langle T^{\mu\nu}(x) T^{\sigma \rho}(x') \rangle = \langle T^{\mu\nu}(x) T^{\sigma \rho}(x') \rangle'_{\rm free} + \langle T^{\mu\nu}(x) T^{\sigma \rho}(x') \rangle'_{\rm int} \, ,
\end{align}
where
\begin{align}
\langle T^{\mu\nu}(x) T^{\sigma \rho}(x') \rangle'_{\rm int}  =  \int_{\mathbb R^d_+} \d^d r \int_{\mathbb R^d_+} \d^d r' \,
\hat t_{\mu\nu}(x,r) \hat t_{\sigma \rho}(x',r') G_\sigma(r,r') \ .
\end{align}

We do not need an explicit form for $G_\sigma(r,r')$ to proceed.  Instead, we recognize the two-point function
\begin{align}
\label{Tsigmarel}
\langle T_{\mu\nu}(x) \sigma(x') \rangle = -N \int_{\mathbb R^d_+} \d^d r \, t_{\mu\nu}(x,r)G_\sigma(r,x') \ .
\end{align}
Conformal symmetry and a Ward identity fix this two point function to have the form \cite{McAvity:1995zd} 
\begin{align}
\langle T_{\mu\nu}(x) \sigma(x') \rangle = - N \frac{2d (4 \mu^2 - 1)}{(d-1) \Omega_{d-1}} \frac{(2z')^{d-2}}{s^{2d}} 
\left( X_\mu X_\nu - \frac{1}{d} \delta_{\mu\nu} \right) v^d \ .
\end{align}
Changing between the hatted $\hat t_{\mu\nu}$ and the unhatted $t_{\mu\nu}$ alters 
$\langle T_{\mu\nu}(x) \sigma(x') \rangle$ by a contact term proportional to $\langle \sigma \rangle \delta(x-x')$, 
as can be seen from (\ref{basic eq for sigma}).  
In fact the two point function $\langle T_{\mu\nu}(x) \sigma(x') \rangle$ more generally 
is arbitrary up to contact terms of this form
\cite{McAvity:1995zd}.   
The stress tensor itself is ambiguous up to a shift $T_{\mu\nu} \to T'_{\mu\nu} = T_{\mu\nu} + c \lambda \sigma \delta_{\mu\nu}$ where $\lambda$ is a position dependent source for $\sigma$ and $c$ is an arbitrary constant.  The stress tensor one point function is untouched when $\lambda = 0$.  Through this shift, however, we can adjust the contact term in the two point function at will.
We choose to regulate the two point function such that $\langle T^\mu_\mu(x) \sigma(x') \rangle = 0$, including 
distributional contributions of the form $\delta(x-x')$.
Through the identification (\ref{Tsigmarel}), we can then be sure that the stress-tensor two-point function
$\langle T^{\mu\nu}(x) T^{\sigma \rho}(x') \rangle$ is traceless.

We can also write $\hat t_{\mu\nu}$ itself in terms of the $X_\mu$.  Inserting the form of $G_\phi$ into the definition (\ref{tmunudef}), we obtain 
\begin{align}
  \frac{1}{N^3} \hat{t}_{\mu \nu} &= \frac{(2z')^2}{s^{2d}} \left(X_\mu X_\nu - \frac{1}{d} \delta_{\mu \nu} \right) f(v) \, , \label{t hat}\\
  f(v) &= - \frac{2}{d-1} \xi (\xi +1) \left((d-2) F(v) \frac{\d}{\d \xi}\left( \xi^2 \frac{\d}{\d \xi} F(v) \right) - d \xi^2 \left(\frac{\d}{\d \xi}F(v) \right)^2  \right) \, . \label{f(v)}
\end{align}
Assembling the pieces, we can write the trace free part of the 
interaction contribution to the stress tensor two point function as 
\begin{align}\label{interaction part}
 \langle T^{\mu\nu}(x) &T^{\sigma \rho}(x') \rangle'_{\rm int} = \\
& N^3 \frac{2d(4\mu^2-1)}{(d-1)\Omega_{d-1}} \int_{{\mathbb R}_+^d} \d^d r \, \left( \frac{2 z \tilde{v}}{\tilde{s}^2 \tilde{s'}^2} \right)^d f(\tilde{v'}) \left(\tilde{X}_\mu \tilde{X}_\nu - \frac{\delta_{\mu \nu} }{d} \right) \left(\tilde{X'}_{\rho} \tilde{X'}_\sigma - \frac{\delta_{\sigma \rho}}{d}  \right) \, ,  \nonumber
\end{align}
where we denote $r = ({\bm r},y)$, $\tilde{s} = (x-r)^2 $, $\tilde{v}^2 = \tilde{s}^2 /(\tilde{s}^2 + 4 z y)$ and $\tilde{s'}^2$, $\tilde{v'}^2$ similarly defined with $x \rightarrow x' $.

To organize the information in the stress-tensor two-point function, we again take advantage of the conformal symmetry.
Tracelessness means the two point function can be characterized by three functions of a cross ratio.  
These functions can be calculated by looking at the special case $x = ({\bf 0}, z)$ and $x' = ({\bf 0}, z')$ and the components \cite{McAvity:1993ue}
\begin{align}
\label{alphadef}
  \langle T_{nn} ({\bm 0},z) T_{nn } ({\bm 0},z') \rangle & = \frac{\alpha(v)}{s^{2d}} \ , \\
  \label{gammadef}
  \langle T_{i n} ({\bm 0},z) T_{k n} ({\bm 0},z') \rangle & = \frac{\gamma(v)}{s^{2d}} \delta_{ik} \ , \\
  \label{epsilondef}
  \langle T_{i j } ({\bm 0},z) T_{ k l } ({\bm 0},z') \rangle & = \frac{\delta(v) \delta_{i j} \delta_{k l } + \epsilon(v) (\delta_{i k}\delta_{j l} + \delta_{i l} \delta_{j k} )  }{s^{2d}}  \ ,
\end{align} 
where by tracelessness, $\alpha = (d-1)((d-1) \delta + 2 \epsilon)$ and we denote the tangential indices as $i, j, \cdots$.  Conservation reduces the information further, to a single function of a cross ratio:
\begin{align}
\label{consrels1}
v \alpha'(v)  -d \alpha(v) &= 2(d-1) \gamma(v) \ , \\
\label{consrels2}
v \gamma'(v) - d \gamma(v) &= \frac{d}{(d-1)^2} \alpha(v) + \frac{(d-2)(d+1)}{d-1} \epsilon(v) \ .
\end{align}
That $\langle T^{\mu\nu}(x) T^{\sigma \rho}(x') \rangle'_{\rm free}$ and $\langle T^{\mu\nu}(x) T^{\sigma \rho}(x') \rangle'_{\rm int}$ are independently traceless means that we can completely specify their form by computing the functions
$\alpha$, $\gamma$ and $\epsilon$ for each structure.  However, they are not independently conserved.  Only the total is conserved.

We first compute $\langle T^{\mu\nu}(x) T^{\sigma \rho}(x') \rangle'_{\rm free}$, restricting to the case $d=3$. 
From the definitions \eqref{alphadef}, \eqref{gammadef}, \eqref{epsilondef}, and plugging in the explicit form of $G_\phi$ into (\ref{free part of TT}), we establish
\begin{align}
  \alpha_{\rm free}(v) &= \frac{N^3 \kappa^2}{9} 
  \frac{(1-v)^{2 \mu -1}}{(1+v)^{2\mu+1}}
 \left\{v \left[9 \mu +v \left(32 \mu ^4 v^2+48 \mu ^3 \left(v^2+1\right) v + 44 v^2+\right. \right. \right. \label{alphafree} \\ 
  &\left. + \left. \left. 4 \mu ^2 \left(9 v^4+8 v^2+9\right)+3 \mu  \left(3 v^4+5 v^2+5\right) v+9 \left(v^2-3\right) v^4-27\right)\right]+9\right\} \ ,  \nonumber  \\
  \gamma_{\rm free}(v) &= -\frac{1}{4}N^3 
   \kappa^2
    \left(\frac{1-v}{1+v} \right)^{2\mu}
  \label{gammafree} \\
  &\quad \times \left\{v \left[6 \mu +v \left(3 v^4+8 \mu ^2 \left(v^2+1\right) +2 \mu  \left(3 v^2-1\right) v-2 v^2+8 \mu ^3 v-2\right)\right]+3\right\} \ ,  \nonumber \\ 
  \epsilon_{\rm free}(v) &= \frac{1}{8} N^3 \kappa^2 
  \frac{(1-v)^{2\mu+1}}{(1+v)^{2\mu-1}}
 \left(v \left(21 \mu +v \left(20 \mu ^2+6 v^2+21 \mu  v+10\right)\right)+6\right) \ .\label{epsilonfree}
\end{align}
One can confirm when $\mu= \pm 1/2$, they reproduce results \cite{McAvity:1993ue} for the free scalar with Dirichlet and Neumann boundary conditions.

Our next task is to calculate the interaction part \eqref{interaction part}. In our setup, with \eqref{bdy block} in $d=3$, $f(v)$ becomes relatively simple:
\begin{align} \label{f(v) in 3d}
  f(v) = \frac{1}{2} \kappa^2 v^3 (v+1)^{-4 \mu } \left(1-v^2\right)^{2 \mu -1} \left(3 \mu +v \left(4 \mu ^2+3 \mu  v+2\right)\right) \, .
\end{align}
The integral in \eqref{interaction part} is generally organized into 
\begin{align}\label{general form of interaction part}
  \int_0^\infty \d y \, \int \d^{d-1} {\bm r} \frac{1}{(2y)^d} \, f_1 (\tilde{\xi}) f_2 (\tilde{\xi}') \left(\tilde{X}_\mu \tilde{X}_\nu - \frac{1}{d} \delta_{\mu \nu} \right) \left(\tilde{X'}_{\rho} \tilde{X'}_\sigma - \frac{1}{d} \delta_{\sigma \rho} \right) \, ,
\end{align}
where we can identify $f_1 = [\xi(\xi + 1)]^{-d/2}$ and $f_2 = N^2 f(v)\xi^{-3}$. 
In Appendix D of \cite{McAvity:1995zd}, the authors investigate a method how to compute \eqref{general form of interaction part} for the case that $f_2 \sim [\xi(\xi + 1)]^{-n}$. We review some aspects of their method in our Appendix \ref{App:conformal integral with boundary} and further generalize it.  The final result is  
\begin{align}\label{hat int}
  \begin{aligned}
    \epsilon_{\rm int}(v) 
    &= c \xi^d 4 \CG''(\xi) \, , \\
  \gamma_{\rm int}(v)
    &= c \xi^d \left[ 4(1+2 \xi)\CG''(\xi) + 8(1+\xi)\xi \frac{\d}{\d \xi} \CG'' \right]  \, ,\\
     \alpha_{\rm int}(v) 
    &= c \xi^d \biggl[ - \frac{8}{d}(d-1)^2 (1+\xi)\xi \left( (2\xi + 1)\frac{\d}{\d \xi} + 2 \right)\CG''(\xi) \\
    & \qquad  \qquad \qquad + \frac{8}{d}(d-1)\CG''(\xi) - \frac{(d-1)^2}{d^3} \Omega_{d-1} f_2(\xi)  \biggr] \, ,
  \end{aligned}
\end{align}
where 
\[
c =N \frac{2d(4\mu^2-1)}{(d-1) \Omega_{d-1}}
\]
is a constant of proportionality.
The function $\CG''(\xi)$ for given $f_2$ is a solution of the second order differential equation (\ref{caliGpp and f2}).

Our strategy for finding $\CG''(\xi)$ is somewhat different than \cite{McAvity:1995zd}.
Rather than pursuing a solution via integral transforms, we solve the differential equation (\ref{caliGpp and f2}).
In $d=3$, defining ${\mathcal F}(v) \equiv {\mathcal G}''(\xi)$, the differential equation takes the form
\begin{align}\label{recasted ode}
  {\mathcal F}''(v) + \frac{2 (3 + 2 v^2)}{v (1-v^2)} {\mathcal F}'(v) + \frac{20}{(1-v^2)^2}{\mathcal F}(v) = S(v) \, ,
\end{align}
where the source term is 
\begin{align}
  S(v) = -\frac{(1+v)^{-2 \mu + 2} (1-v)^{2 \mu + 2} (3 \mu + v(2 + 3 v \mu + 4 \mu^2))}{96 \pi v^5} \ .
\end{align}
The expression \eqref{recasted ode} has two homogeneous solutions:
\begin{align}
\label{homone}
  {\mathcal F}_1(v) &= \frac{(1-v^2)^5}{v^5} \ , \\
  \label{homtwo}
{\mathcal F}_2(v) &= \frac{-3v + 14 v^3 - 14 v^7 + 3 v^9 + 3(1-v^2)^5 \tanh^{-1}(v)}{128 v^5} \ ,
\end{align}
with Wronskian
\begin{align}
  {\mathcal W} = {\mathcal F}_1(v) {\mathcal F}_2'(v) - {\mathcal F}_1'(v) {\mathcal F}_2(v)  = \frac{(1-v^2)^5}{v^6} \ .
\end{align}
Our boundary conditions are that ${\mathcal F}(v)$ is less singular than $v^{-5}$ in the coincident $v\to 0 $ limit and
vanishes faster than $(v-1)$ in the boundary $v \to 1$ limit, leading to the solution of interest
\begin{align}
\label{Fsolfinal}
 {\mathcal F}(v) =  -{\mathcal F}_2 \int_v^1 \frac{\mathcal F_1(v') S(v')}{{\mathcal W}(v')} \d v' - {\mathcal F}_1 \int_0^v \frac{\mathcal F_2(v') S(v')}{{\mathcal W}(v')} \d v' \ .
\end{align}
These boundary conditions are consistent with the behavior of the integral (\ref{relint}) in the 
$v \to 0$ and $v \to 1$ limits.

As the two point function satisfies a conservation Ward identity, all of the information in the two point function is encoded in the single function $\alpha(v)$.  With a solution for ${\mathcal F}(v)$ in hand, we can plug it into (\ref{hat int}) to obtain $\alpha_{\rm int}(v)$ and add to that the ``free'' contribution $\alpha_{\rm free}(v)$ (\ref{alphafree}) to obtain the net result.  
The remaining functions $\gamma(v)$ and 
$\epsilon(v)$ can then be constructed from the conservation relation (\ref{consrels1}) and (\ref{consrels2}).  
Alternatively and as a cross check, one can obtain $\gamma(v)$ and $\epsilon(v)$ from (\ref{hat int}), (\ref{gammafree}) and (\ref{epsilonfree}).  The result is the same.

\begin{figure}
  \centering 
  \includegraphics[width=12cm]{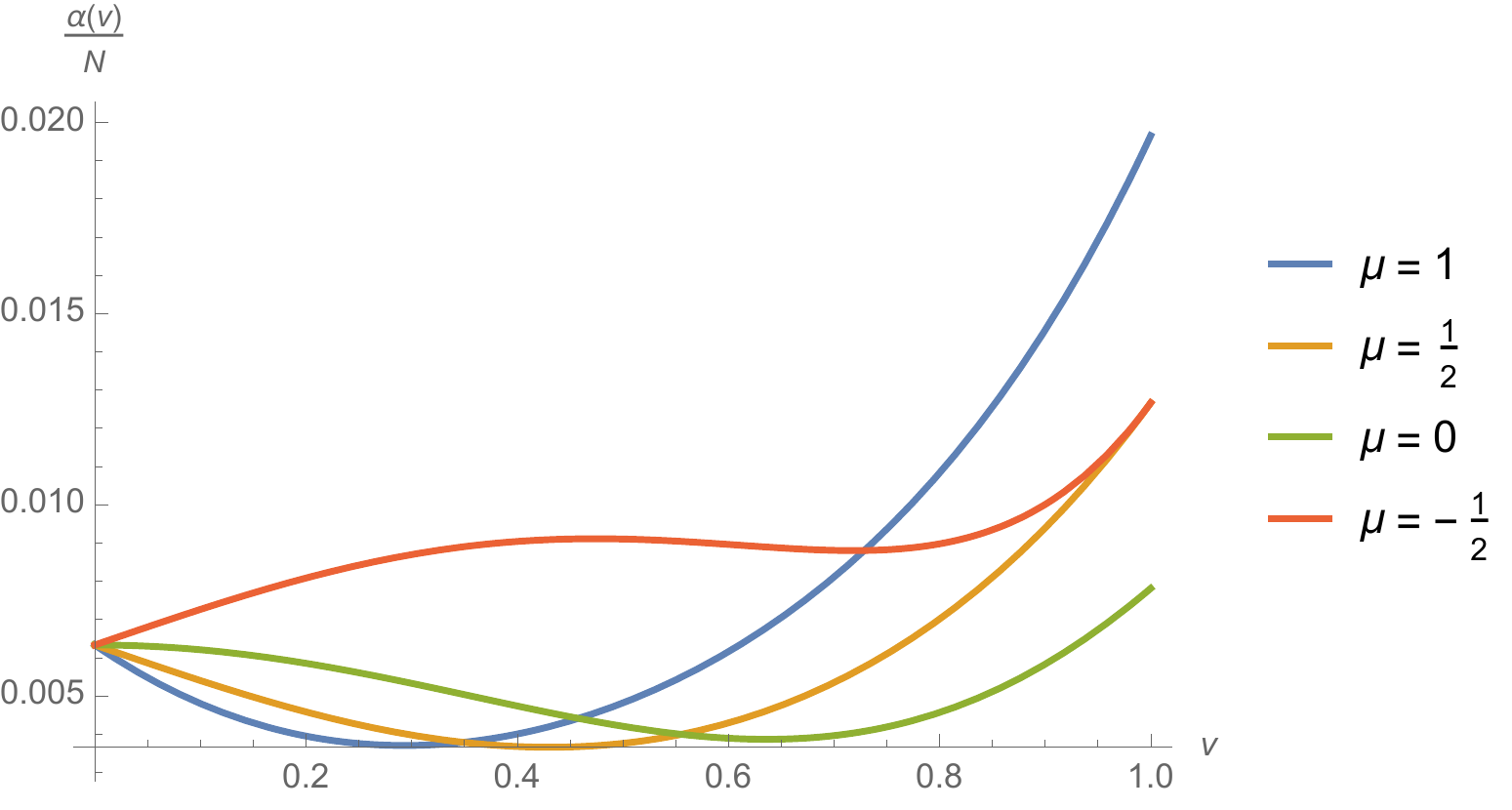}
  \caption{We plot $\alpha(v)$ for various values of $\mu$. All curves start with the same value at $v=0$ while they end with different values at $v=1$. } 
  \label{fig:alphas}
\end{figure}

We have not been able to find a closed form expression for the integral (\ref{Fsolfinal}), but nevertheless, this presentation of the solution is very convenient.  
We will use it to analyze the limits $\alpha(0)$ and $\alpha(1)$ next.  In the subsections to come, we present closed form expressions in four special cases $\mu = \pm \frac{1}{2}$, 0, and 1.  Figure \ref{fig:alphas} presents a graph of $\alpha(v)$ in these four cases.
Finally in section \ref{sec:boundary decomposition}, we decompose $\alpha(v)$ into 
bulk and boundary conformal blocks for general $\mu$, 
which will give us some information about the spectrum of bulk and boundary conformal primaries in this theory.

The value of $\alpha(v)$ in the coincident limit is universal, $\alpha(0) = N/ 16 \pi^2$ regardless of $\mu$.  
The interaction part $\alpha_{\rm int} (0)$ vanishes, and the answer is given just by the free part $\alpha_{\rm free}(0)$, 
which is equal to $N / 16 \pi^2$. 
Without a boundary, the two point function is fixed up to not just a function but a constant.  In the coincident limit of
our theory with a boundary,
we expect to recover this constant, or central charge, $\alpha(0)$, also sometimes called $C_T$.  
This number should be independent of boundary conditions.  Here we find it is also independent of 
the quasi-marginal coupling $g$.

On the other hand, $\alpha(1)$ is very sensitive to $\mu$ and through $\mu$, to the coupling $g$.
 It is known that $\alpha(1)$ gives the normalization of the displacement operator two-point function and thus is also related to a boundary central charge in the trace anomaly \cite{Herzog:2017xha}, a fact whose consequences 
 we will investigate in section \ref{sec:disc}. 
It is straightforward to analyze 
   \begin{equation}
   \label{alphaone}
    \alpha(1) = -\frac{64 (4\mu^2-1)N }{\pi} \int_0^1 \frac{{\mathcal F}_2(v) S(v)}{{\mathcal W}(v)} \d v \, ,
    \end{equation}
numerically for $\mu>1/2$ and also via saddlepoint approximation in the large $\mu$ limit.  With a little bit of effort,
we can extend the region of validity of this formula to $\mu > -1$ through a minimal subtraction procedure, removing
the power law divergences at the upper range of the integral $v \to 1$.  Beyond $\mu=-1$ (the unitarity bound for the boundary operators), the subtraction procedure
becomes ambiguous because of the presence of logarithms.

\begin{figure}
  \centering 
  a) \includegraphics[width=7cm]{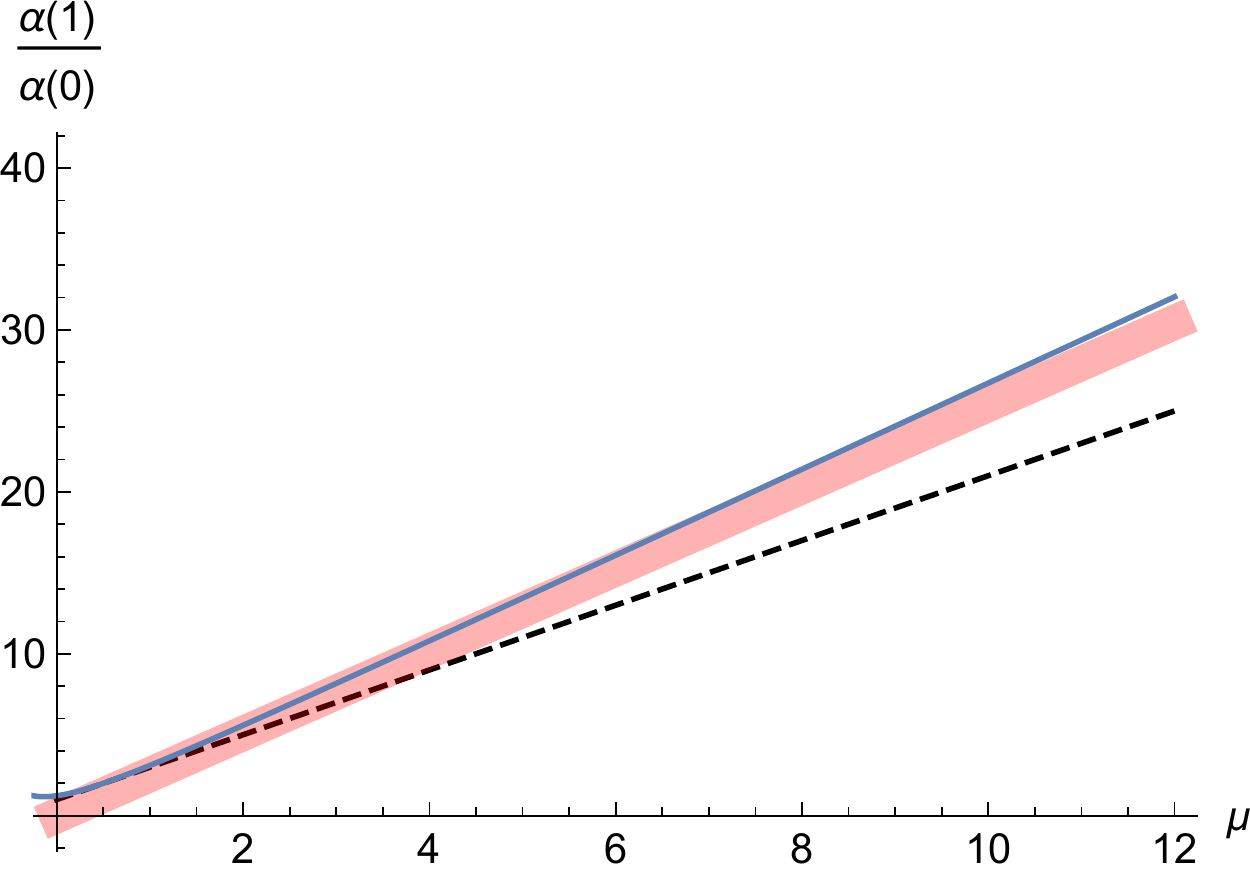} \; \ 
  b) \includegraphics[width=7cm]{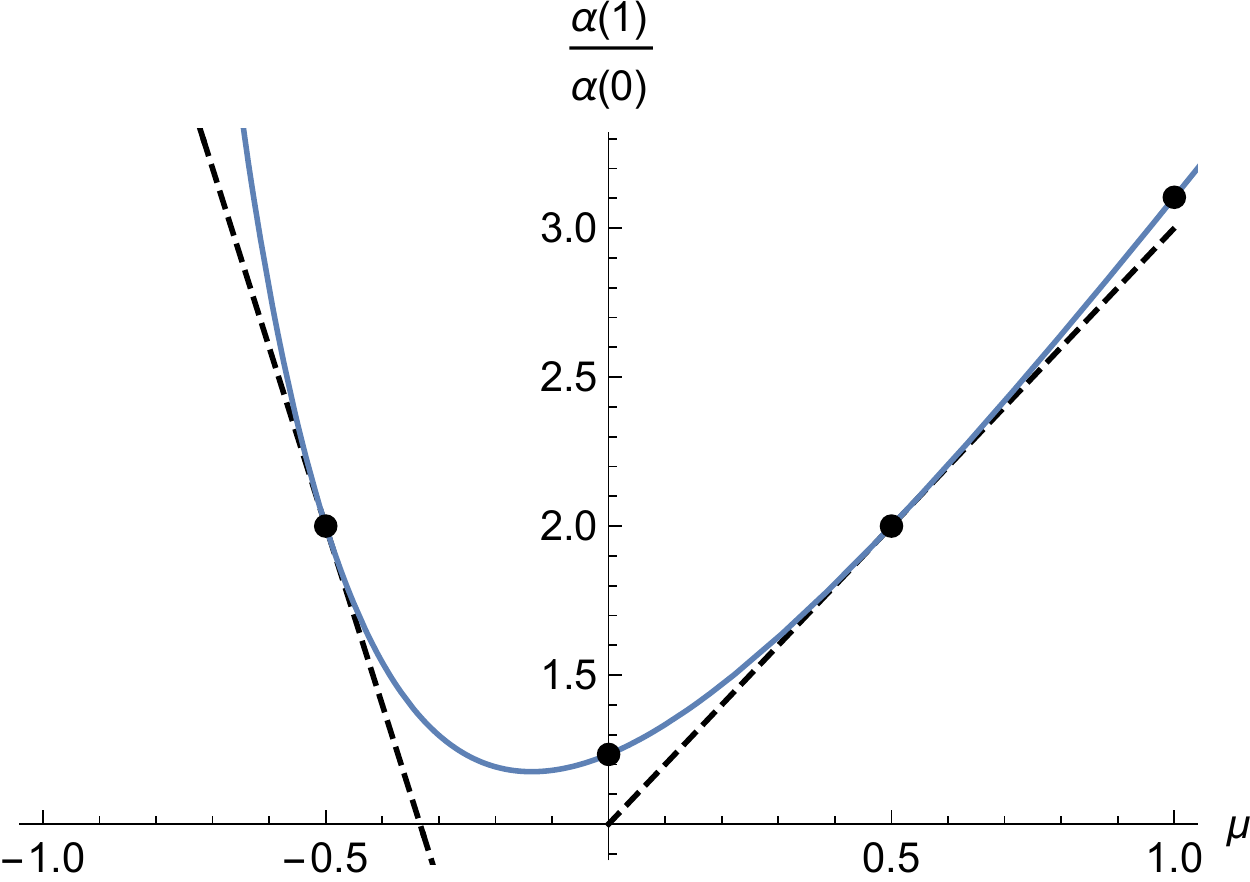}
  \caption{
  A plot of $\alpha(1)/\alpha(0)$ vs.\ $\mu$.  The solid blue line was computed numerically.  
  The dashed black lines are tangents at $\mu = \pm 1/2$.  The thick red line is the large 
  $\mu$ saddle point approximation.  The black dots are analytically computed points. (b) zooms in on the small $\mu$ region of (a).  There is a minimum at approximately $\mu = -0.136$.
  } 
  \label{fig:numericalpha}
\end{figure}

  We provide plots of $\alpha(1)$ in figure \ref{fig:numericalpha}.
  The saddlepoint approximation yields
  \begin{align}
  \frac{\alpha(1)}{\alpha(0)} \sim \mu \frac{8}{15} e^{\frac{1-\sqrt{13}}{2}} \sqrt{\pi \left(50 + \frac{172}{\sqrt{13}} \right)}  \sim 2.54 \mu \ .
  \end{align}
Numerically, we see that for $g<0$ (equivalently $-\frac{1}{2} < \mu < \frac{1}{2}$), 
$\alpha(1)$ satisfies the inequality 
$\alpha(1) < 2 \alpha(0)$ while for the coupling in the domain
$0 < g < 16 \pi^2$ (equivalently $|\mu| > \frac{1}{2}$), we have instead $\alpha(1) > 2 \alpha(0)$.
  It is unclear to us whether the $g<0$ cases are physical.  On the one hand, they correspond
to an unbounded $\phi^6$ potential.  On the other, from the point of view of a Weyl equivalent hyperbolic space, the curvature at the maximum of the potential is above the BF bound.   
In ref.\ \cite{Herzog:2017xha}, it was found that
$\alpha(1) < 2 \alpha(0)$ in the case of a theory with interactions confined to the boundary.  Thus our ``less physical'' case agrees with the previous study.
Interestingly, 
the $\alpha(v)$ we find for the $\mu=0$ case is the same as that found in \cite{McAvity:1995zd} 
for $d=3$ $\phi^4$ theory at large $N$ with Dirichlet boundary conditions.

\subsection{Perturbative expansion by small coupling}
We begin with the small coupling limit. Recalling $\mu^2 = (4-g/4\pi^2)^{-1}$, in the small $g$ limit, $\mu$ can be expanded as
\begin{align}
  \mu =  \pm \left( \frac{1}{2}+\frac{ g }{64 \pi ^2}\right) +O\left(g^2 \right) \, .
\end{align}
To leading order, we are allowed to set $\mu = \pm 1/2$ in $f(v)$ due to the overall coefficient in the interaction part \eqref{interaction part}. In these cases, we find 
\begin{align}
  S(v) &= \mp \frac{(1-v^2)^3}{64 \pi v^5} \ .
\end{align}
Enforcing the boundary conditions described above, we find a solution that in fact vanishes 
at $v=0$ and $1$. We obtain 
\begin{align}
  {\mathcal F} (v) &= \mp \frac{(1-v)^5 (1+v) (v (3 + v (8 + 3 v)) - 3 (1+v)^4 \tanh^{-1}(v))}{1536 \pi v^5} \ . 
\end{align}
Writing ${\mathcal F}(v) = {\mathcal G}''(\xi)$ in terms of $\xi$ and using the relations \eqref{hat int}, the interaction parts are given as follows:
\begin{align}
  \alpha_{\rm int}(v) & = \pm \frac{ g N}{64 \pi ^2 } \frac{  v \left(\left(9 v^4+6 v^2+9\right) \tanh ^{-1}(v)+v ((4-9 v) v-9)\right)}{48 \pi ^2} + O(g^2)  \, , \\
  \gamma_{\rm int}(v) &=  \pm \frac{ g N }{64 \pi ^2 } \frac{  (v-1) v \left(3 (v+1)^2 \left(v^2+1\right) \tanh ^{-1}(v)-v (v (3 v+4)+3)\right)}{32 \pi ^2 (v+1)} + O(g^2)\, , \\
  \epsilon_{\rm int}(v) &=  \pm  \frac{ g N}{64 \pi ^2 } \frac{ (v-1)^2 v \left(3 (v+1)^4 \tanh ^{-1}(v)-v (v (3 v+8)+3)\right)}{128 \pi ^2 (v+1)^2} + O(g^2) \, .
\end{align}
To combine with the free part, we also expand \eqref{alphafree}-\eqref{epsilonfree} in the small coupling limit. 
 The net result for $\alpha(v)$ is
\begin{align}
 \alpha(v) =& \, N\left( \frac{1+v^6}{16 \pi^2} + \frac{g}{512 \pi^4} v \left(v+v^3 + 3(1-v^2)^2 \tanh^{-1}(v) \right)  \right) \nonumber \\
& \pm N \Biggl( - \frac{3v (1-v^2)^2}{32 \pi^2}  + \frac{g}{1024 \pi^4} 
\biggl(v (1 + (v-3)v)(1+v^2) + \nonumber \\
& \qquad \; \; \; \;  +  (-4 + 3v + 2 v^3 + 3 v^5 - 4 v^6 ) \tanh^{-1} (v) \biggr) \Biggr) + O(g^2) \ , 
\end{align}
where the plus sign corresponds to Dirichlet boundary conditions and the minus sign to Neumann.
As the total result satisfies the conservation Ward identities (\ref{consrels1}) and (\ref{consrels2}), we can easily construct 
$\gamma(v)$ and $\epsilon(v)$ from $\alpha(v)$. 

The boundary limit of $\alpha(v)$ is interesting because it represents the normalization of the displacement operator two point function. We find
\begin{align}
  \alpha(1) &= \frac{N}{8 \pi^2} \left(1+ \frac{ 2g \mp g }{64 \pi ^2}  \right) + O(g^2) \ ,
\end{align}
which suggests that $\alpha(1)$ starts as an increasing function of the coupling $g$. We also see that the bulk limit of $\alpha$ is $\alpha(0) = N / 16 \pi^2$, which implies that $\alpha(1) > 2\alpha(0)$ when $g>0$.  

\subsection{$\mu=0$: strong coupling limit}
The next example is the $\mu=0$ case, which corresponds to the $g \rightarrow -\infty$ limit. It is not clear that
the theory is stable in this limit, as the $\phi^6$ potential is unbounded below.  We can nevertheless naively proceed with the same analysis of the stress tensor two point function.
In this case, we have
\begin{align}
  S(v) &= - \frac{(1-v^2)^2}{48 \pi v^4} \ .
\end{align}
We discover a solution 
\begin{align}
  \begin{aligned}
    {\mathcal F}(v) &= \frac{1-v^2}{6144 \pi v^5} \left\{ 6 (1-v^2)^4 \tanh^{-1}(v) \log(v) \right.\\
    & + 2v (3 - v^2 + v^4 - 3 v^6 - (1+v^2)(3 - 14 v^2 + 3 v^4) \log(v) ) \\
    &\left. + 3(1-v^2)^4 ( \Li_2(-v) - \Li_2(v) ) \right\} \, ,
  \end{aligned}
\end{align}
where $\Li_n (x)$ is a polylogarithm. 
This solution scales as $v^{-3}$ in the coincident limit and $(1-v)^4$ in the boundary limit.

Adding $\alpha_{\rm int}$ and $\alpha_{\rm free}$ together, 
the information in the stress tensor two point function is encapsulated in the single function
\begin{align}
    \alpha (v) & = \frac{N}{512 \pi ^2} \left\{v \left(3 v^4+2 v^2+3\right) \left(4 \text{Li}_2(v)-\text{Li}_2\left(v^2\right)\right) \right. \\
    & \left. +4 \left(8-8 v^6+19 v^4-19 v^2+v \log (v) \left(3 \left(v^3+v\right)-\left(3 v^4+2 v^2+3\right) \tanh ^{-1}(v)\right)\right) \right\} \, , \nonumber
\end{align}
from which we may construct $\gamma(v)$ and $\epsilon(v)$ using the conservation equations
(\ref{consrels1}) and (\ref{consrels2}).
We observe that 
\begin{align}
  \alpha (1) = \frac{N}{128} \, .
\end{align}
As usual we find $\alpha(0) = N/16\pi^2$, and  so it follows $\alpha(1) < 2\alpha(0)$.
While the inequality $\alpha(1) < 2\alpha(0)$ is consistent with results that were found for a theory with only boundary interactions \cite{Herzog:2017xha}, it is not clear that the $\mu=0$ case studied here is physical -- because of the unbounded potential. 

As mentioned already, 
this case was studied in \cite{McAvity:1995zd}.  There, the authors computed the two-point function of the stress tensor in $\phi^4$ theory at large $N$, for general dimension $d \leq 4$.  In the particular case $d=3$ with ``Dirichlet'' boundary conditions, their two point function reduces to ours.  Their answer, valid for general $d$, is written in term of a hypergeometric function
$_3 F_2$.  With some effort, one can demonstrate that in fact the two solutions are the same at $d=3$.

\subsection{One more special case: $\mu=1$}
For $\mu = 1$, we have 
\begin{align}
  S (v) = - \frac{(1-v)^4(1+v)^2}{32 \pi v^5} \, ,
\end{align}
and find that 
\begin{align}
    \CF (v) &= - \frac{1-v^2}{6144 \pi v^5} \Biggl( - 2 v (1-v^2) (-9 + v (16 + v(-6 + v (-32 + v(-9 + 16 v))))) 
    \nonumber \\
    & + 6(-3v + 11 v^3 + 11 v^5 - 3 v^7 + 3 (1-v^2)^4 \tanh^{-1}(v)) \log(v) \nonumber \\
    & + 9 (1-v^2)^4 (\Li_2(-v) - \Li_2(v)) \Biggr) \ ,
\end{align}
with the same boundary conditions as before.  This function diverges as $v^{-3}$ in the coincident limit and vanishes as
$(v-1)^5$ in the boundary limit.
Inserting the result for ${\mathcal F}(v)$ into (\ref{hat int}) and adding the free result, we obtain
\begin{align}
     \alpha (v) & = -\frac{N}{{256 \pi ^2 }} \left\{9 v \left(3 v^4+2 v^2+3\right) (\text{Li}_2(-v)-\text{Li}_2(v)) \right. \\
    &  +18 v \log (v) \left(\left(3 v^4+2 v^2+3\right) \tanh ^{-1}(v)-3 \left(v^3+v\right)\right) \nonumber  \\
    &\left. +2 v (v (v (v (8 v (v+3)+21)+16)-21)+24)-16 \right\} \,  ,  \nonumber
\end{align}
from which we may construct $\gamma(v)$ and $\epsilon(v)$ using the conservation relations
(\ref{consrels1}) and (\ref{consrels2}).
Taking the boundary and bulk limit, we end up with 
\begin{align}
  \alpha(1) = N \left(\frac{9}{128}-\frac{1}{2 \pi ^2} \right) \, , \quad \alpha(0) = \frac{N}{16\pi^2} \, , 
\end{align}
which implies $\alpha(1) > 2 \alpha(0)$ in the case at hand $\mu=1$, 
and $\alpha(1)|_{\mu = 1} > \alpha(1)|_{\mu = \frac{1}{2}}$.

\section{Conformal block decomposition} \label{sec:boundary decomposition}
So far we have calculated two-point functions of the conserved current and the stress tensor. 
By using the operator product expansion, we can re-express these correlation functions as sums over
exchanged operators.  Given conformal symmetry, these sums naturally arrange themselves
into conformal blocks, where each block compactly represents the exchange of a conformal primary operator
and all its descendants.  There are two natural limits: the coincident limit in which the sum is over 
bulk scalar primary operators and the boundary limit in which case the sum is over boundary primaries.
See \cite{Liendo:2012hy,Herzog:2017xha} for a lengthier discussion of these issues.

As a warm up, consider the
 two-point function of a scalar operator $\CO$ with dimension $\Delta$.  In the boundary limit, the two point function is decomposed as follows
\begin{align}
  \langle \CO_\Delta (x_1) \CO_\Delta (x_2) \rangle = \frac{1}{|x_1 -x_2|^{2 \Delta}}\, \xi^\Delta \left(a_{\mathcal O}^2  + \sum_{\Delta'} \bryOPE_{\Delta'}^2 G_{\textrm{bry}}(\Delta', v) \right) \, ,
\end{align}
where 
\begin{align}
  G_{\mathrm{bry}} (\Delta,v) = \xi^{-\Delta} \, _2F_1 \left(\Delta,1-\frac{d}{2}+\Delta, 2-d+2\Delta, - \frac{1}{\xi} \right)  \, ,
\end{align}
and $a_\CO$ is the coefficient in the one-point function of $\CO$. The 
$\bryOPE_{\Delta'}$ are proportional to boundary OPE coefficients and the $\Delta'$ are the dimensions of exchanged boundary operators. 
In contrast, in the coincident limit, we can decompose the two-point function into a sum over bulk conformal blocks
\begin{align}
  \langle \CO_\Delta (x_1) \CO_\Delta (x_2) \rangle = \frac{1}{|x_1 - x_2|^{2 \Delta}}  \left( \lambda + \sum_{\Delta' \neq 0} 
  a_{\Delta'} \bulkOPE_{\Delta'} G_{\rm bulk}(\Delta', v) \right) \ ,
  \end{align}
where
\begin{align}
G_{\rm bulk}(\Delta, v) = \xi^{\frac{\Delta}{2}} {}_2 F_1 \left( \frac{\Delta}{2}, \frac{\Delta}{2}, 1 - \frac{d}{2} + \Delta; -\xi \right) \ .
\end{align}
The $a_{\Delta}$ are the coefficients in the one point functions of the scalar operators that appear in the OPE of ${\mathcal O}(x)$ with itself, while the $\bulkOPE_\Delta$ are the usual OPE coefficients.  There may be an identity operator in this OPE, whose contribution to the bulk conformal block expansion we denote by $\lambda$.

Our task is to extend this decomposition to spinning operators and to 
determine the $\bryOPE_{\Delta}^2$, $a_{\Delta} \bulkOPE_{\Delta}$, and scaling dimensions $\Delta$ in the theory.  

\subsection{Boundary conformal block decomposition}

\subsubsection*{The conserved current}
In the case of $\langle J_\mu (x_1) J_\nu (x_2) \rangle $,  we will focus on the decomposition of $\pi(v)$. The decomposition of $C(v)$ follows from the current conservation Ward identity (\ref{JJWard}). 

The boundary block expansion for $\pi(v)$ is given by 
\begin{align} \label{bry expansion of pi}
  \pi (v) = \xi^{d-1} \left(\bryOPE^2_{(0)} \pi^{(0)}_{\mathrm{bry}} (v) + \sum_{\Delta \geq d-2} \bryOPE^2_{\Delta} \pi^{(1)}_{\mathrm{bry}} (\Delta, v) \right) \, ,
\end{align}
where the indices (0) and (1) denote the spins of the exchanged operators.
According to \cite{Herzog:2017xha}, we have
\begin{align}
  \pi^{(0)}_{\mathrm{bry}} (v) = \frac{1}{2} (v^{-1} - v)^{d-1} (v^{-1} + v) \, ,
\end{align}
and the spin one conformal blocks are 
\begin{align}
  \pi_{\rm bry}^{(1)}(\Delta, v) = \xi^{-\Delta-1} {}_2 F_1 \left( 1 + \Delta, 1- \frac{d}{2} + \Delta, 2-d+2 \Delta; -\frac{1}{\xi} \right) \ .
\end{align}
In order to fix $\bryOPE_\Delta^2$, we need to expand \eqref{pi} and the right hand side of \eqref{bry expansion of pi} around $v=1$ and compare them term by term. For our $d=3$, large $N$ case, we find that
\begin{align}
  (1-v)^{2+2\mu} (1+v)^{2-2 \mu} \xi^{-2} =  \sum_{j=0}^\infty \frac{(1+j)(1+j+2 \mu)}{2^{4(\mu+j)} (1 + 2j + 2 \mu)} \pi_{\rm bry}^{(1)}(3 + 2 \mu + 2 j, v) \ .
\end{align}
Note that the left hand side transforms under $v \to 1/v$ with a phase factor $(-1)^{4 + 2 \mu}$.  The boundary blocks on the other hand transform with phase factor $(-1)^{\Delta + 1}$, which rules out a contribution from boundary blocks with dimension $2 \mu$ plus an even number.  

In general, we find that the current two point function involves exchanging a tower of spin one boundary conformal primaries with dimensions $\Delta = 3 + 2\mu + 2j$, for $j$ a non-negative integer.
This spectrum of dimensions is natural if we can associate the boundary limit of the field $\phi_\alpha$ 
with an operator ${\mathcal O}_\alpha$ 
of dimension $\mu+1$.  The operators in the tower should then have the schematic form
$\Box^j (\partial_\mu {\mathcal O}_{[\alpha})  ({\mathcal O}_{\beta]})$.

It is useful to analyze the free cases, 
 $\mu = \pm \frac{1}{2}$, in a little more detail.  
 In the Neumann case $\mu = -\frac{1}{2}$, the boundary limit of the field $\phi_\alpha$ does indeed have dimension 
 $\Delta = \frac{1}{2}$.   On the other hand $\mu = \frac{1}{2}$ corresponds to Dirichlet boundary conditions, 
 in which case it is most natural to think of the operator at the bottom of the tower as the boundary limit of 
 $\partial_n \phi_\alpha$ instead of $\phi_\alpha$ itself.

\subsubsection*{Stress tensor}
Next we consider the boundary decomposition of $\alpha(v)$ in $\langle T_{\mu \nu} (x_1) T_{\sigma \rho} (x_2) \rangle$. The decomposition has the form
\begin{align}
  \alpha(v) = \xi^d \left(\bryOPE^2_{(0)} \alpha^{(0)}_{\mathrm{bry}} (v) + \sum_{\Delta \geq d-1} \bryOPE^2_{\Delta} \alpha^{(2)}_{\mathrm{bry}} (\Delta, v)  \right) \, ,
\end{align}
with 
\begin{align}
  \alpha^{(0)}_{\rm bry} (v) & = \frac{1}{4(d-1)}(v^{-1} - v)^d (d (v^{-1} + v)^2 - 4) \, , \\
  \alpha_{\rm bry}^{(2)}(\Delta, v) &= \xi^{-\Delta-2} {}_2 F_1 \left( 2 + \Delta, 1 - \frac{d}{2} + \Delta, 2-d+2 \Delta; -\frac{1}{\xi} \right) \ .
\end{align}
Here $\alpha^{(0)}_{\rm bry}$ is a conformal block corresponding to the displacement operator, i.e.\ a scalar operator conjugate to the location of the boundary.  No other scalar operators contribute.  
The $\alpha_{\rm bry}^{(2)}(\Delta, v)$ are spin two boundary operators with scaling dimension $\Delta$.  There is no spin one contribution to the decomposition. 

Before giving the general solution, let us recall what happens in the free case $g=0$ \cite{Liendo:2012hy,Herzog:2017xha}.  
 In the free theory, 
one can make use of the following identity
\begin{align}
  \frac{1}{2}\left(1+v^{2 d}\right)=\xi^{d}\left(\alpha_{\mathrm{\text{bry}}}^{(0)}(v)+\sum_{j \in 2 \mathbb{Z}^{*}} \bryOPE_{j}^{2} \alpha_{\mathrm{\text{bry}}}^{(2)}(d+j, v)\right) \, ,
\end{align}
where ${\mathbb Z}^*$ is the set of non-negative integers and 
\begin{align}
  \bryOPE_{j}^{2}=\frac{2^{-d-2 j} \sqrt{\pi} \Gamma(d+j-1) \Gamma(d+j+2)}{\Gamma(d) \Gamma\left(\frac{d}{2}-1\right) \Gamma(j+3) \Gamma\left(\frac{d+1}{2}+j\right)} \, .
\end{align}
The full result is then obtained by tweaking the series representation of $\frac{1}{2}(1+v^{2d})$ slightly.
For Dirichlet conditions $\alpha_{\rm bry}^{(2)}(d,v)$ is removed while for Neumann conditions, its contribution is doubled.

While we cannot find a general closed form solution for $\alpha(v)$, it is straightforward to expand (\ref{Fsolfinal}) near $v=1$ and from this integral representation, construct the first few terms in a series expansion for $\alpha(v)$ near the boundary.

We find the dimensions of the 
spin-two boundary blocks are $4+2 \mu + 2j$ where $j$ is a non-negative integer and $\alpha(v)$ is expanded as 
\begin{align}\label{proposed form of decomposition}
  \alpha (v) =  \xi^3 \left(\alpha (1) \alpha^{(0)}(v) + \sum_{j =0}^\infty \bryOPE_j^2 \alpha^{(2)}_{\rm bry}(4+2\mu+2j,v) \right) \, .
\end{align}
The pattern here is similar to that of the current-current two-point function boundary decomposition.  
If there is an operator ${\mathcal O}_\alpha$
with dimension $\mu + 1$ corresponding to the boundary limit of $\phi_\alpha (x)$, then we find
 spin-two operators of the form $\Box^j (\partial_\mu  {\mathcal O}_\alpha ) ( \partial_\nu {\mathcal O}_\alpha)$ with scaling dimension of $4 + 2\mu + 2 j$. 
The first few coefficients in this sum are %
  \begin{equation}
  \begin{array}{|c|c|c|c|c|c|}
  \hline
   & \multicolumn{5}{c|}{16 \pi^2 N^{-1} \bryOPE_j^2 } \\
   \hline
   j &  \mu & \mu = -\frac{1}{2} & \mu = 0 & \mu = \frac{1}{2} & \mu = 1 \\
   \hline
   0 &\frac{(1+\mu)(2+\mu)}{2^{4\mu-2}(3 + 2 \mu)^2} & 3 & \frac{8}{9} & \frac{15}{64} & \frac{3}{50}  \\
   \hline
   1 & \frac{3(1+\mu)(3+\mu)}{2^{4\mu+2} (5 + 2\mu)^2} & \frac{15}{64} & \frac{9}{100} & \frac{7}{256} & \frac{3}{392} \\
   \hline
   2 & \frac{3(2+\mu)(4+\mu)(3+2\mu)}{2^{4\mu+5}(5+2\mu) (7 + 2 \mu)^2} & \frac{7}{256} & \frac{9}{980} & \frac{45}{16384} & \frac{25}{32{,}256} \\
   \hline
   3 & \frac{5 (2+\mu)(5+\mu)(5+2\mu)}{2^{4\mu+9}(7+2\mu)(9+2\mu)^2} &\frac{45}{16{,}384} & 
   \frac{125}{145{,}152} & \frac{33}{131{,}072} & \frac{35}{495{,}616}  \\
   \hline
   4 & \frac{15 (3+\mu)(6+\mu)(5+2\mu)}{2^{4\mu+14} (9+2\mu)(11+2\mu)^2} & 
   \frac{33}{131{,}072} & \frac{75}{991{,}232} & \frac{91}{4{,}194{,}304} & \frac{735}{121{,}831{,}424} \\
   \hline
  \end{array}
    \end{equation}
    while $\alpha(1)$ was given in (\ref{alphaone}).  The $\mu = \pm \frac{1}{2}$ columns agree with the $1+v^{2d}$ decomposition discussed above.

\subsection{Bulk block decomposition}
Now let us switch gears and dicuss the bulk decomposition. Unlike the boundary decomposition, it is not necessary for $a_\Delta c_\Delta$ to be positive. In fact we will see many of these coefficients are negative. In addition, the bulk primaries exchanged with the boundary can only be scalars since the one-point functions of spinning operators vanish due to conformal symetry. 
\subsubsection*{Conserved current}

For the conserved current, to keep expressions simpler, it is useful to decompose $D(v)$ into conformal blocks rather than $\pi(v)$.  The decomposition takes the general form \cite{Liendo:2012hy,Herzog:2017xha}
\begin{align}
D(v) = \sum_{\Delta \neq 0} a_\Delta \bulkOPE_\Delta D_{\rm bulk}(\Delta, v) \ ,
\end{align}
where
\begin{align}
D_{\rm bulk}(\Delta, v) = \xi^{\frac{\Delta}{2}} {}_2 F_1 
\left( 1 + \frac{\Delta}{2}, 1 + \frac{\Delta}{2}, 1- \frac{d}{2} + \Delta; -\xi \right) (1 + \xi) \ .
\end{align}
For us, the sum over $\Delta$ is restricted to positive integers.  The first few coefficients are as follows:
\begin{equation}
\begin{array}{|c|c|c|c|c|c|}
\hline
 & \multicolumn{5}{|c|}{16 \pi^2  \, a_\Delta c_\Delta} \\
 \hline
 \Delta & {\rm general} \, \mu & \mu = -\frac{1}{2} & \mu = 0 & \mu = \frac{1}{2} & \mu = 1 \\
 \hline
1 & -4 \mu & 2 & 0 & -2&  -4 \\
2 & 4(4 \mu^2-1) & 0 & -4 & 0 & 12 \\
3 &  -8 \mu(4 \mu^2 -1 ) & 0 & 0 & 0 & - 24\\
4 & \frac{4}{3} (32 \mu^4 - 24 \mu^2 + 1) & -4 & \frac{4}{3}& -4 & 12 \\
5 & -\frac{32}{3} \mu (\mu^2 - 1)(4\mu^2-1)& 0 & 0 & 0 & 0 \\
6 & \frac{16}{105} \left(224 \mu^6 - 400 \mu^4 + 167 \mu^2 - 9\right)& \frac{12}{7} & -\frac{48}{35} 
& \frac{12}{7} & -\frac{96}{35} \\
\hline
\end{array} \ .
\end{equation}
Numerically, we have observed some patterns associated with these coefficients.  
The $a_\Delta c_\Delta$ are polynomials of degree $\Delta$ in $\mu$, without a definite sign.  However, the coefficient of the $\mu^\Delta$ term in the polynomial has sign $(-1)^j$.  Thus for large enough $\mu$, the coefficients $a_\Delta c_\Delta$ should have alternating sign.  
Another interesting feature of this decomposition is that for odd $\Delta>3$, the polynomial 
coefficients have a factor $\mu (\mu^2-1)(4\mu^2 -1)$.  Thus, they will vanish in the $\mu = \pm \frac{1}{2}$, 0, and 1 cases.

\subsubsection*{Stress tensor}

For the stress tensor, the bulk conformal block decomposition is simpler for the function
\begin{align}
A(v) = \frac{d^2}{(d-1)^2} \alpha(v) + 4 \gamma(v) + \frac{2(d-2)}{d-1} \epsilon(v) \ ,
\end{align}
than for $\alpha(v)$.
For $A(v)$, we find \cite{Liendo:2012hy,Herzog:2017xha} 
\begin{align}
A(v) = \sum_{\Delta \neq 0} a_\Delta \bulkOPE_\Delta A_{\rm bulk} (\Delta, v) \ ,
\end{align}
where
\begin{align}
A_{\rm bulk}(\Delta, v) = \xi^{\frac{\Delta}{2}} {}_2 F_1 
\left( 2 + \frac{\Delta}{2}, 2 + \frac{\Delta}{2}, 1- \frac{d}{2} + \Delta; -\xi \right) (1 + \xi)^2 \ .
\end{align}
The first few coefficients are 
\begin{equation}
\begin{array}{|c|c|c|c|c|c|}
\hline
& \multicolumn{5}{|c|}{16 \pi^2 N^{-1} \, a_\Delta \bulkOPE_\Delta}\\
\hline
\Delta & \mu & \mu = -\frac{1}{2} & \mu = 0 & \mu = \frac{1}{2}& \mu = 1 \\
\hline
1 & - \frac{9\mu}{8}
&\frac{9}{16}  & 0 & -\frac{9}{16} & -\frac{9}{8} \\
2 &
2(4\mu^2-1) & 0
& - 2 & 0 & 6 \\
3 & -5\mu(4 \mu^2-1)& 0 &  0 & 0 & - 15 \\
4 & 2(4\mu^2-1)^2 & 0 & 2 & 0 & 18 \\
5 & -\frac{7}{3} \mu (4\mu^2-1)^2 & 0 & 0 & 0 & -21 \\
6 & a_6 \bulkOPE_6 &
6 & -\frac{16}{525} (29 + 105 \log(v))& 6 & 
\frac{16}{175} (123 - 315 \log v) \\
\hline
\end{array}
\end{equation}
where
\begin{equation}
a_6 \bulkOPE_6 = N \frac{-29 + 1632 \mu^2 - 3474 \mu^4 + 2240 \mu^6 - 105 (4\mu^2-1)^2 \log(v)}{525 \pi^2} \ .
\end{equation}
Similar to the bulk block decomposition for $\langle J_\mu(x_1) J_\nu(x_2) \rangle$, the bulk block decomposition here is again over scalar operators with positive integer dimension.

We can see that for general $\mu$ the bulk blocks with dimension $\Delta  \geq 6$ have logs in their expansion.
The appearance of a logarithm is a problem as it introduces a scale to what is supposed to be a scale 
invariant theory.
We can gain some insight from the $\mu=0$ case, where our expression matches a result from
\cite{McAvity:1995zd}.  In this older paper, the authors computed the stress tensor two-point function for
$\phi^4$ theory in a large $N$ limit and general dimension.  In the specific case $d=3$, their expression
matches ours, and so we see that that their conformal block expansion must also involve logarithms.
Using their result to move away from $d=3$, there is a scalar operator of dimension $2d$ and a second of dimension 6 that contribute to the conformal block decomposition.  The coefficients of these conformal blocks are equal and opposite in the $d\to 3$ limit and scale as $1 / (d-3)$.  The collision and mixing of these two operators in $d=3$ produces the logarithm.\footnote{We would like to thank H.~Osborn for discussion on this point.  For readers interested in duplicating the result, there is a typo in (5.34) \cite{McAvity:1995zd}.  A factor of $v^d$ multiplying a ${}_2 F_1$ hypergeometric function should be $v^{2d}$.}
A similar degeneracy happens in integer dimensions $d >4$ for operators of dimension $2d$, 
but not in $d=4$ where the theory is free.
It is interesting that the lack of positivity of the bulk conformal block expansion allows these two diverging coefficients 
to cancel.
We would like to explore how this mixing is affected by $1/N$ corrections although it is important to note that in our context at least, there may be a problem that the theory is no longer conformal at subleading order in $1/N$.
(A similar log in a one point function was pointed out in \cite{Herzog:2019bom}, where it was likely related to an 
anomaly in the trace of the stress tensor.)

  \section{Discussion}
  \label{sec:disc}
  
  One of the motivations for this work was to look for tractable examples of boundary CFT where the trace anomaly coefficients $a$ and $b$ in (\ref{traceanomaly}) could be computed.  
  These quantities are thus far known only in a few examples.  One is the conformally coupled scalar.
There are two types of Weyl invariant boundary conditions: Dirichlet and Robin.  The central charges for these two choices are 
$a^{(D)} = -\frac{1}{96}$ \cite{Nozaki:2012qd},
$a^{(R)} = \frac{1}{96}$ \cite{Jensen:2015swa},
and
$b^{(D)} = b^{(R)} = \frac{1}{64}$ \cite{Fursaev:2016inw}.  The Robin boundary condition involves an extrinsic curvature, and for a planar boundary reduces to the Neumann condition.

These two quantities $a$ and $b$ are easily computable for our $\phi^6$ theory.  
The charge $a$ can be extracted from the effective action of the theory on hyperbolic space $H_3$.  
We have already computed the potential density $V$ in section \ref{Sec:O(N)atlargeN} (see figure \ref{fig:solutions}).
The effective action at leading order in $N$ is the integral of $V$ over $H_3$, and since $V$ is constant, we have $W = V \, \Vol(H_3)$.  
Take a line element on $H_3$ of the form 
$\d s^2 = L^2 [ \d \tau^2 + \sinh^2 \tau ( \d \theta^2 + \sin^2 \theta d \phi^2) ]$ where $L$ is the radius of curvature and
the coordinates satisfy $\tau > 0$, $0 \leq \theta \leq \pi$, and $0 \leq \phi \leq 2 \pi$ with the conformal boundary at $\tau \to \infty$.  
Of course the volume of this metric is formally infinite, but we can regularize by cutting off the integration 
``close to the boundary'' 
at 
$e^{\tau_{\rm max}} =  L \Lambda$.  We find that
\begin{align}
\Vol(H_3) = - 2 \pi \log (L \Lambda) \ .
\end{align}
The stress tensor trace can be re-interpreted as a scale variation of the partition function, $Z = e^{-W}$.  
For hyperbolic space with this $S^2$ boundary, we find then 
$\Lambda \partial_\Lambda W = -2 a \log L \Lambda$ and 
\begin{equation}
\label{apiV}
a = \pi V \ .
\end{equation}
Reassuringly, we find that in the free Neumann and Dirichlet cases ($g=0$), we recover the free field results
$a= \pm N \frac{1}{96}$.  More generally, for nonzero $g$ and $\phi_\alpha = 0$, we find
the simple scaling $a = -N \frac{\mu}{48}$.  The result for the extraordinary boundary condition can be read off from 
figure \ref{fig:solutions}.
By the monotonicity theorem for $a$ \cite{Jensen:2015swa}, 
a boundary renormalization group flow can only take one from a larger value of $V$ to a smaller one.
If we insist on boundary unitarity ($\mu > -1$), then we see from figure \ref{fig:stability} 
that $a < \frac{N}{48}$ is bounded above.

The other coefficient $b$ can be extracted from the displacement two point function.  As the displacement operator is the boundary limit of the $T^{nn}$ component of the stress tensor, we can also extract $b$ from the stress tensor two point function.  From \cite{Herzog:2017kkj}, we have
\begin{equation}
\label{balphaone}
b = \frac{\pi^2}{8} \alpha(1) \ .
\end{equation}

With these results (\ref{apiV}) and (\ref{balphaone}) in hand, we can check that a pair of conjectures about these coefficients $a$ and $b$ appears to be false.  One could perhaps object that our counter example is not a good one -- that our theory is only conformal in the strict large $N$ limit.
Nevertheless, we feel that the failure of the conjectures in this case gives evidence that the conjectures are likely incorrect.
  
In ref.\ \cite{Herzog:2017kkj}, it was posited that $a$ could be extracted from the stress tensor two point function, 
in particular
\begin{equation}
a = \frac{\pi^2}{9} \left( \epsilon(1) - \frac{3}{4} \alpha(1) + 3C \right) \, ,
\end{equation}
where $C$ is the central charge of a decoupled 2d CFT living on the boundary.  In our case, there is no such decoupled CFT and $C$ vanishes.  Moreover, $\epsilon(1)$ vanishes except in the special cases $\mu = \pm \frac{1}{2}$. 
Thus the conjecture boils down to the statement that $a = - \pi^2 \alpha(1) / 12$, which is manifestly
not true in the disordered case, comparing the actual result $a = -\mu N / 48$ with figure \ref{fig:numericalpha}, which is 
not linear in $\mu$. 
Thus the conjecture appears to be wrong.

Another conjecture, this time concerning $b$ and $\alpha(1)$, was discussed in ref.\ \cite{Herzog:2017xha}.  
The authors speculated that perhaps $\alpha(1)$ was bounded above by $2 \alpha(0)$ because that is 
what they observed in a graphene like theory where the interaction was confined to the boundary.  
The value $\alpha(0)$
is related to the coincident limit of the stress tensor two point function.  From figure \ref{fig:numericalpha}, it is clear
that this bound is satisfied only in the range $| \mu | < \frac{1}{2}$, or equivalently $g<0$.  
For $g>0$, on the other hand, $\alpha(1) > 2 \alpha(0)$.  

\vskip 0.1in

We leave many interesting questions unanswered in this work.  How do $1/N$ corrections change the story?  What happens if we look in $3-\epsilon$ dimensions?  Can we say more about the classically marginal boundary term
$(\vec \phi^2)^2$ in the case where the relevant boundary term $\vec \phi^2$ is tuned to zero?  Is there more that can be said about the logarithms that appear in the bulk conformal block expansion of the stress tensor two point function?  
Are there any interesting experimental systems that are described by our large $N$ model?  We hope to return to some of these topics in the future.

  \section*{Acknowledgments}
  We would like to thank Itamar Shamir and Abhay Shrestha for discussion as well as Kuo-Wei Huang, Hugh Osborn and Hans Diehl for correspondence.  We would also like to thank Julio Virrueta for collaboration during the early stages of this project.
  C.H.\ was supported in part by the U.K.\ Science \& Technology Facilities Council Grant ST/P000258/1 and by a Wolfson Fellowship from the Royal Society.
  N.K.\ was supported in part by the Program for Leading Graduate Schools, MEXT, Japan and by JSPS Research Fellowship for Young Scientists. N.K\ was also supported by World Premier International Research Center Initiative (WPI Initiative), MEXT, Japan.

\appendix
\section{Conformal integral with boundary} \label{App:conformal integral with boundary}
In this appendix, we review the method to compute the integral \eqref{general form of interaction part}, which was studied in Appendix D of \cite{McAvity:1995zd}. Let us start with the following integral,
\begin{align}
  f(\xi)= & \int_{0}^{\infty} \mathrm{d} z \int \mathrm{d}^{d-1} \mathbf{r} \frac{1}{(2 z)^{d}} f_{1}(\tilde{\xi}) f_{2}(\tilde{\xi}^{\prime})\  , 
  \\
  \; \; & \tilde{\xi} = \frac{(x-r)^2}{4yz} , \quad \tilde{\xi}' = \frac{(x'-r)^2}{4y'z} , \quad r = ({\bm r},z) \, . \nonumber
\end{align}
To obtain the form of $f(\xi)$, we consider the problem backwards and perform the following invertible integral transform,
\begin{align}\label{basic conformal integral with boundary}
  \hat{f}(\rho) &= \frac{1}{(4y y')^g} \int \d^{d-1} {\bm x} \, f(\xi) \, \\
& = \frac{\pi^g}{\Gamma(g)} \int_0^\infty \d u \, u^{g-1} f(u+\rho) \, , \nonumber
\end{align}
where $\rho = (y-y')^2/(4y y')$ and $g = (d-1)/2$. The inverse transform is given as follows,
\begin{align}\label{inverse transform}
  f(\xi) = \frac{1}{\pi^g \Gamma(-g)} \int_0^\infty \d \rho \, \rho^{-g -1} \hat{f} (\rho + \xi) \, .
\end{align}
Employing the above transform, \eqref{basic conformal integral with boundary} can be recast as 
\begin{align} \label{transformed conformal integral}
  \hat{f}(\rho)=\int_{0}^{\infty} \mathrm{d} z \frac{1}{2 z} \hat{f}_{1}(\tilde{\rho}) \hat{f}_{2}\left(\tilde{\rho}^{\prime}\right) \ , \quad \tilde{\rho}=\frac{(y-z)^{2}}{4 y z} \ , \quad \tilde{\rho}^{\prime}=\frac{\left(y^{\prime}-z\right)^{2}}{4 y^{\prime} z} \, .
\end{align}
Then if we can compute $\hat{f}(\rho)$ by \eqref{transformed conformal integral}, it enables us to obtain $f(\xi$) by the inverse integral transform. To this end, we first change variables $z = e^{2 \theta}$, $y = e^{2 \theta_1}$ and $y' = e^{2 \theta_2}$. \eqref{transformed conformal integral} becomes
\begin{align}\label{transformed conformal integral-2}
  \hat{f}\left(\sinh ^{2}\left(\theta_{1}-\theta_{2}\right)\right)=\int_{-\infty}^{\infty} \mathrm{d} \theta \hat{f}_{1}\left(\sinh ^{2}\left(\theta-\theta_{1}\right)\right) \hat{f}_{2}\left(\sinh ^{2}\left(\theta-\theta_{2}\right)\right) \, .
\end{align}
Taking the Fourier transform of \eqref{transformed conformal integral-2},
\begin{align}
  \tilde{\hat{f}}(k) = \int_{-\infty}^{\infty} \d \theta \, e^{i k \theta} \hat{f}(\sinh^2 \theta) \, ,
\end{align}
the convolution property gives us the following simple relation,
\begin{align}
  \tilde{\hat{f}}(k)=\tilde{\hat{f}}_{1}(k) \tilde{\hat{f}}_{2}(k) \, ,
\end{align}
which makes it possible to compute $f(\xi)$ from the given $f_1 (\xi)$ and $f_2 (\xi)$. One strategy is to perform the series of integral transforms that converts $f_i$ to $\tilde{\hat{f_i}}$ and then to use the convolution property to obtain $\tilde{\hat{f}}$. Performing an inverse Fourier transform and \eqref{inverse transform}, we obtain $f(\xi)$ in the end. The success of the method depends highly on the form of $f_i$, but for some of 
the $f_i$ of interest, we can do these integral transforms.

The spin structures add another layer of complexity to the evaluation of \eqref{general form of interaction part}. Let us introduce the differential operator
\begin{align}
  \tilde{\mathcal{D}}_{\mu \nu} \equiv \partial_{\mu} \partial_{\nu}+\frac{1}{y}\left(n_{\mu} \partial_{\nu}+n_{\nu} \partial_{\mu}\right)-\frac{1}{d} \delta_{\mu \nu}\left(\partial^{2}+\frac{2}{y} n \cdot \partial\right) \, .
\end{align}
This operator $\tilde{\CD}_{\mu \nu} $ allows us to re-express the $X_\mu X_\nu - \frac{\delta_{\mu\nu}}{d}$ tensor structure in terms of derivatives acting on a function of a cross ratio:
\begin{align}
  \tilde{\mathcal{D}}_{\mu \nu} \mathcal{F}(\xi)=\frac{1}{z^{2}}\left(X_{\mu} X_{\nu}-\frac{1}{d} \delta_{\mu \nu}\right) \xi(1+\xi) \mathcal{F}^{\prime \prime}(\xi) \, ,
\end{align}
which allows us to write \eqref{general form of interaction part} as
\begin{align}\label{cariG acting tildeDs}
  \CG_{\mu \nu \sigma \rho} = (4 z z')^2 \tilde{\CD}_{\mu \nu} \tilde{\CD}'_{\sigma \rho} \CG(\xi) \, ,
\end{align}
where for $i=1$ or 2
\begin{align}
\label{relint}
  \CG(\xi) = \int_0^\infty \d y \, \int \d^{d-1} {\bm r} \, \frac{1}{(2y)^d} \CF_1(\tilde{\xi}) \CF_2 (\tilde{\xi}') \ , \quad f_i(\xi)=4 \xi(1+\xi) \mathcal{F}_i^{\prime \prime}(\xi) \, .
\end{align}
In the above expression we can play the same game not for $f_i$, but for $\CF_i$. The point is that we don't need to transform $\CF$ itself because using integration by parts we can write
\begin{align} \label{carif transform}
  \hat{\CF}(\rho) = \frac{\pi^g}{\Gamma(g+2)} \int_0^\infty \d u \, u^{g+1} \CF'' (u +\rho) \ .
\end{align}
In our setup, $f_1(\xi) = (\xi (1+\xi))^{-d/2}$ and corresponding integral transforms are easily done by 
\begin{align}
  \hat{\mathcal{F}}_{1}\left(\sinh ^{2} \theta\right)=\frac{1}{2} S_{d} \frac{1}{d(d+1)} e^{-(d+1)|\theta|}, \quad \tilde{\hat{\mathcal{F}}}_{1}(k)=\frac{1}{d} S_{d} \frac{1}{k^{2}+(d+1)^{2}} \, .
\end{align} 
Now suppose we know $\tilde{\hat{\CF}}_2$.  Then we have
\begin{align}
  { \hat \CG(\sinh^2(\theta))} = \frac{S_d}{d} \frac{1}{2\pi} \int \d \theta \, e^{- i k \theta} \, \frac{1}{(d+1)^2 + k^2} \tilde{\hat{\CF}}_2 (k) \, ,
\end{align}
from which we find that 
\begin{align}
  \left( (d+1)^2 - \frac{\d^2}{\d^2 \theta}\right) \hat{\CG} (\sinh^2(\theta)) = \frac{S_d}{d} \hat{\CF}_2 (\sinh^2(\theta)) \, .
\end{align}
Using the integral transform (\ref{basic conformal integral with boundary}), 
we can pull this differential equation back to one involving ${\mathcal G}(\xi)$
\begin{align}
  \left(\xi(1+\xi) \frac{\mathrm{d}^{2}}{\mathrm{d} \xi^{2}}+d\left(\xi+\frac{1}{2}\right) \frac{\mathrm{d}}{\mathrm{d} \xi}-d\right) \mathcal{G}(\xi)=-\frac{1}{4 d} S_{d} \mathcal{F}_2(\xi) \, ,
\end{align}
or equivalently ${\mathcal G}''(\xi)$,
\begin{align} \label{caliGpp and f2}
  \left(\xi(1+\xi) \frac{\mathrm{d}^{2}}{\mathrm{d} \xi^{2}}+(d+4)\left(\xi+\frac{1}{2}\right) \frac{\mathrm{d}}{\mathrm{d} \xi}+d+2\right) \mathcal{G}^{\prime \prime}(\xi)=-\frac{1}{d} S_{d} \frac{1}{16 \xi(1+\xi)} f_2(\xi) \, .
\end{align}

\bibliographystyle{JHEP}
\bibliography{phi6_with_boundarynew}

\end{document}